\documentclass[aps,prd,preprint,draft,showpacs,eqsecnum]{revtex4}
\usepackage{graphicx,epsf,amssymb}
\newcommand{\be}{\begin{equation}} 
\newcommand{\ee}{\end{equation}}

\newcommand{\bea}{\begin{eqnarray}} 
\newcommand{\eea}{\end{eqnarray}} 
\newcommand{\Tr}{{\rm Tr}}

\newcommand{\NeqFour}{{\cal N} =4}

\def\coeff#1#2{\relax{\textstyle {#1 \over #2}}\displaystyle}

\newif\ifdraft
\drafttrue
\newif\ifpreprint
\preprinttrue

\def\sect#1{section~{\ref{#1}}}
\def\app#1{appendix~{\ref{#1}}}

\def\fig#1{fig.~{\ref{#1}}}

\def\Tr{\, {\rm Tr}}

\def\NeqFour{{\cal N}=4}

\def\ns{n_{\mskip-2mu s}}
\def\nf{n_{\mskip-2mu f}}

\def\spa#1.#2{\left\langle#1\,#2\right\rangle}
\def\spb#1.#2{\left[#1\,#2\right]}
\def\sand#1.#2.#3{%
\left\langle\smash{#1}{\vphantom1}^{-}\right|{#2}%
\left|\smash{#3}{\vphantom1}^{-}\right\rangle}
\def\sandp#1.#2.#3{%
\left\langle\smash{#1}{\vphantom1}^{-}\right|{#2}%
\left|\smash{#3}{\vphantom1}^{+}\right\rangle}
\def\sandpp#1.#2.#3{%
\left\langle\smash{#1}{\vphantom1}^{+}\right|{#2}%
\left|\smash{#3}{\vphantom1}^{+}\right\rangle}
\def\sandpm#1.#2.#3{%
\left\langle\smash{#1}{\vphantom1}^{+}\right|{#2}%
\left|\smash{#3}{\vphantom1}^{-}\right\rangle}
\def\sandmp#1.#2.#3{%
\left\langle\smash{#1}{\vphantom1}^{-}\right|{#2}%
\left|\smash{#3}{\vphantom1}^{+}\right\rangle}
\def\sandmm#1.#2.#3{%
\left\langle\smash{#1}{\vphantom1}^{-}\right|{#2}%
\left|\smash{#3}{\vphantom1}^{-}\right\rangle}
\def\spab#1.#2.#3{\sandmm#1.#2.#3}
\def\spba#1.#2.#3{\sandpp#1.#2.#3}
\def\spaa#1.#2.#3.#4{\sandmp#1.{#2#3}.#4}
\def\spbb#1.#2.#3.#4{\sandpm#1.{#2#3}.#4}

\newbox\charbox
\newbox\slabox
\def\s#1{{      
        \setbox\charbox=\hbox{$#1$}
        \setbox\slabox=\hbox{$/$}
        \dimen\charbox=\ht\slabox
        \advance\dimen\charbox by -\dp\slabox
        \advance\dimen\charbox by -\ht\charbox
        \advance\dimen\charbox by \dp\charbox
        \divide\dimen\charbox by 2
        \raise-\dimen\charbox\hbox to \wd\charbox{\hss/\hss}
        \llap{$#1$}
}}

\def\eqn#1{eq.~(\ref{#1})}
\def\Eqn#1{Equation~(\ref{#1})}
\def\eqns#1#2{eqs.~(\ref{#1}) and~(\ref{#2})}

\def\qb{{\bar q}}

\def\ib{{\bar\imath}}
\def\e{\epsilon}

\def\Gr{{\rm Gr}}

\def\sign{{\mathop{\rm sign}\nolimits}}
\def\lr{\leftrightarrow}

\def\Split{\mathop{\rm Split}\nolimits}
\def\tree{{(0)}}
\def\oneloop{{(1)}}

\def\twoloop{{(2)}}
\def\SUSY{{\rm SUSY}}

\def\Ord{{\cal O}}

\def\Kh{{\hat K}}

\def\si{\sigma}
\def\sandp#1.#2.#3{%
\left\langle\smash{#1}{\vphantom1}^{+}\right|{#2}%
\left|\smash{#3}{\vphantom1}^{+}\right\rangle}
\def\ksl{\s{k}}
\def\Ksl{\s{K}}

\def\Soft{{\cal S}}

\def\Res{\mathop{\rm Res}}
\def\tlambda{{\tilde\lambda}}
\def\psl{\s{p}}
\def\onehalf{1/2}
\def\f{{\! f}}

\newbox\ourfigbox
\def\SizedFigureWithCaption#1#2#3{%
\setbox\ourfigbox
  \hbox{\hss\epsfxsize #1 \epsfbox{#2}\hss}
\hbox to \wd\ourfigbox{\vbox{\noindent\copy\ourfigbox\break
\vskip -6mm      \hbox to \wd\ourfigbox{\hss#3\hss}}}
}

\def\llongrightarrow{%
\relbar\mskip-0.5mu\joinrel\mskip-0.5mu\relbar
     \mskip-0.5mu\joinrel\longrightarrow}
\def\inlimit^#1{\buildrel#1\over\llongrightarrow}
\def\dash{\hbox{-\kern-.02em}}

\def\spa#1.#2{\left\langle#1\,#2\right\rangle}
\def\spb#1.#2{\left[#1\,#2\right]}
\def\spash#1.#2{\vphantom{\hat K}\spa{\smash{#1}}.{\smash{#2}}}
\def\spbsh#1.#2{\vphantom{\hat K}\spb{\smash{#1}}.{\smash{#2}}}
\def\lor#1.#2{\left(#1\,#2\right)}
\def\sand#1.#2.#3{%
\left\langle\smash{#1}{\vphantom1}^{-}\right|{#2}%
\left|\smash{#3}{\vphantom1}^{-}\right\rangle}
\def\sandpp#1.#2.#3{%
\left\langle\smash{#1}{\vphantom1}^{+}\right|{#2}%
\left|\smash{#3}{\vphantom1}^{+}\right\rangle}
\def\sandpm#1.#2.#3{%
\left\langle\smash{#1}{\vphantom1}^{+}\right|{#2}%
\left|\smash{#3}{\vphantom1}^{-}\right\rangle}
\def\sandmp#1.#2.#3{%
\left\langle\smash{#1}{\vphantom1}^{-}\right|{#2}%
\left|\smash{#3}{\vphantom1}^{+}\right\rangle}

\begin{document}
\hfuzz 10 pt


\ifpreprint
\noindent
UCLA/05/TEP/15
\hfill SLAC--PUB--11134
\hfill Saclay/SPhT--T05/058
\hfill hep-ph/0505055
\fi

\title{The Last of the Finite Loop Amplitudes in QCD%
\footnote{Research supported in part by the US Department of 
 Energy under contracts DE--FG03--91ER40662 and DE--AC02--76SF00515}}

\author{Zvi Bern}
\affiliation{ Department of Physics and Astronomy, UCLA\\
\hbox{Los Angeles, CA 90095--1547, USA}
}

\author{Lance J. Dixon} 
\affiliation{ Stanford Linear Accelerator Center \\ 
              Stanford University\\
             Stanford, CA 94309, USA
}

\author{David A. Kosower} 
\affiliation{Service de Physique Th\'eorique\footnote{Laboratory 
   of the {\it Direction des Sciences de la Mati\`ere\/}
   of the {\it Commissariat \`a l'Energie Atomique\/} of France.}, 
   CEA--Saclay\\ 
          F--91191 Gif-sur-Yvette cedex, France
}

\date{May 4, 2005}

\begin{abstract}
We use on-shell recursion relations to determine the one-loop QCD
scattering amplitudes with a massless external quark pair and an
arbitrary number ($n-2$) of positive-helicity gluons.  These
amplitudes are the last of the unknown infrared- and
ultraviolet-finite loop amplitudes of QCD.  The recursion relations
are similar to ones applied at tree level, but contain new non-trivial
features corresponding to poles present for complex momentum arguments
but absent for real momenta.  We present the relations and the compact
solutions to them, valid for all $n$.  We also present compact forms
for the previously-computed one-loop $n$-gluon amplitudes with a
single negative helicity and the rest positive helicity.
\end{abstract}

\pacs{11.15.Bt, 11.25.Db, 11.25.Tq, 11.55.Bq, 12.38.Bx \hspace{1cm}}

\maketitle



\renewcommand{\thefootnote}{\arabic{footnote}}
\setcounter{footnote}{0}


\section{Introduction}
\label{IntroSection}

The computation of new one-loop amplitudes in perturbative gauge
theories, and in QCD in particular, will be a prerequisite for
theoretical studies related to the experimental program at CERN's
upcoming Large Hadron Collider (LHC).  The discovery and study of new
physics beyond the standard $SU(3)\times SU(2)\times U(1)$ model of
particle interactions will depend on our ability to calculate
higher-order corrections to a wide variety of processes in its
component gauge theories.  Computations of tree-level scattering
amplitudes are a first but insufficient step.  The size and
scale-variation of the strong coupling constant imply that a basic
quantitative understanding must also include the one-loop amplitudes
which enter into next-to-leading order corrections to cross sections.
Such corrections are also required to build a theoretical base for a
program of precision measurements at hadron
colliders~\cite{GloverReview}.  Precision measurements at the SLAC
Linear Collider (SLC) and CERN's Large Electron Positron (LEP)
collider have proven the power of such a program in advancing our
understanding of short-distance physics.

Recent years have seen important progress in this theoretical program;
yet a wide and seemingly hostile province still severs us from our
goal, encouraging us to seek additional tools for performing the
required calculations.  The last year in particular has seen new
progress in the computation of
tree-level~\cite{CSW,Higgs,Currents,RSVNewTree,
BCFRecurrence,BCFW,LuoWen,BFRSV,BadgerMassive} and
one-loop~\cite{BST,BCF7,NeqFourSevenPoint,BCFII,NeqFourNMHV,OtherGaugeCalcs}
gauge-theory amplitudes stimulated by Witten's
proposal~\cite{WittenTopologicalString} of a {\it weak--weak\/}
coupling duality between $\NeqFour$ supersymmetric gauge theory and
the topological open-string $B$ model in twistor space, generalizing
Nair's earlier description~\cite{Nair} of the simplest gauge-theory
amplitudes.  Further investigations along these lines have revealed
new aspects of the underlying twistor structure of gauge
theory~\cite{RSV,Gukov,BBK,CSWII,CSWIII,BBKR,CachazoAnomaly}.  For
a recent review, see ref.~\cite{CSReview}.

Recursion relations, originally introduced by Berends and
Giele~\cite{BGRecurrence}, have long been recognized as an efficient
and elegant method for computing tree-level amplitudes.  Other related
approaches~\cite{OtherNumerical}, as well as various computer-driven
approaches such as MADGRAPH~\cite{Madgraph}, have also been employed.
Recently, Britto, Cachazo and Feng wrote down~\cite{BCFRecurrence}
recursion relations, employing only on-shell amplitudes (at complex
values of the external momenta).  These relations were
stimulated by the compact forms of seven- and higher-point tree
amplitudes~\cite{NeqFourSevenPoint,NeqFourNMHV,RSVNewTree} that
emerged from infrared consistency equations~\cite{UniversalIR}.  They
were shown to yield compact expressions for 
next-to-maximally-helicity-violating (NMHV) tree
amplitudes~\cite{LuoWen}.  The same authors and Witten gave a
simple and elegant proof~\cite{BCFW} of the relation using
special complex continuations of the external momenta.  

The proof, which we review in \sect{RecursionReviewSection}, is actually
quite general, and applies to any rational function of the external
spinors satisfying certain scaling and factorization properties. 
Indeed, it has since been applied to amplitudes with
massive particles~\cite{BadgerMassive}, and in gravity as 
well~\cite{GravityRecurrence}.
The generality of the proof
suggested that it should be useful for finding on-shell recursion
relations at one loop.  We previously wrote down~\cite{OnShellRecurrenceI}
such relations for the (infrared and ultraviolet) finite $n$-gluon 
amplitudes in QCD, ${\cal A}_n^{\oneloop}(1^\pm,2^+,3^+,\ldots,n^+)$,
for which all gluons (or all but one) have the same helicity.
Unlike the situation for massless tree amplitudes, in an application
to general loop amplitudes in a non-supersymmetric theory,
factorization in complex momenta is qualitatively different from
that in real momenta.  Accordingly, we had to address new issues,
in particular the appearance of double poles in the complex analytic
continuation.

In this paper, we examine another application of such on-shell relations,
to one-loop $n$-point amplitudes with one pair of massless external quarks
and $(n-2)$ positive-helicity gluons,  
${\cal A}_n^{\oneloop}(1_{\bar{q}}^-,2_q^+,3^+,\ldots,n^+)$.
This set of helicity amplitudes, together with the above $n$-gluon amplitudes 
and their partners under parity, are zero identically at tree level 
due to supersymmetry Ward identities~\cite{Susy}.  This is because 
massless quarks differ from gluinos only in color manipulations 
which are essentially trivial at tree level.  
At one loop, the difference in color factors
between quarks and gluinos permit the amplitudes to be non-vanishing.
However, any infrared and ultraviolet divergences would have to be
proportional to the corresponding tree amplitude, which vanishes.
Hence this set of helicity amplitudes is finite.
Because all the ``zeroes'' have been filled in at one loop,
none of the corresponding two-loop helicity amplitudes can be
finite.  For example, the first two-loop four-gluon scattering
amplitude to be computed~\cite{TwoLooppppp}, 
${\cal A}_4^{\twoloop}(1^+,2^+,3^+,4^+)$, has infrared
divergences similar to a typical one-loop amplitude.
Thus the amplitudes we compute in this paper represent the last
finite loop amplitudes to be computed in massless QCD.

We calculated the five-point amplitude, ${\cal
A}_n^{\oneloop}(1_{\bar{q}}^-,2_q^+,3^+,4^+,5^+)$ (together with all
the other helicity configurations) long ago~\cite{TwoQuarkThreeGluon},
but no other results in this class of amplitudes exist in pure QCD.
On the other hand, related QED and mixed QED/QCD amplitudes, 
containing a massless external $e^+e^-$ pair and arbitrarily many 
positive-helicity photons or gluons, were computed by 
Mahlon~\cite{MahlonQED,Mahlon}. 

In the recursive approach to the finite quark-gluon amplitudes that we take, we
must treat a new type of pole in addition to the double poles
encountered in the pure-gluon case~\cite{OnShellRecurrenceI}.  These
are poles where the collinear limit in real momenta is not singular,
yet nonetheless a single pole arises for complex momenta.  We shall call these
``unreal'' poles. An understanding of these unreal poles is essential
for constructing correct recursion relations.  In this paper we will
not provide a first principles derivation of the unreal poles, but
will instead take a pragmatic approach.  First we determine the unreal pole
contributions to the recursion relation by reconstructing the known 
five-point amplitudes~\cite{TwoQuarkThreeGluon}.  Then we use our
determination of the unreal poles at five points to write down
a pair of recursion relations for an arbitrary number of external 
legs.  
We find compact solutions to the two relations, 
valid for all $n$.
For a subset of the amplitudes, we also find a set of recursion relations
in which unreal poles are absent.

To confirm these relations we perform non-trivial consistency 
checks of the factorization properties of the solutions. 
As an additional cross check, we use our QCD amplitudes to compute 
mixed QED/QCD and pure QED amplitudes, by carrying out sums over 
appropriate permutations of the gluon momenta.  The permutation sums 
are designed to remove non-abelian self-couplings, thus allowing a 
quark-anti-quark pair to mimic an electron-positron pair.  We then 
compare the permutation sums with Mahlon's earlier results for the 
same amplitudes~\cite{MahlonQED,Mahlon}.

As a side benefit of our analysis we also find a compact form for the
previously computed~\cite{Mahlon} one-loop $n$-gluon amplitudes with a
single negative helicity and the rest positive.  This form is obtained
from the quark amplitudes by taking the momenta of the quark and 
anti-quark to be collinear, and making use of the 
known~\cite{Neq4Oneloop,BernChalmers,OneloopSplit}
factorization properties.

On-shell recursion relations may provide a technique, 
complementary to the unitarity-based method of Dunbar and the 
authors~\cite{Neq4Oneloop,UnitarityMachinery,TwoLooppppp} 
(and its more recent refinements~\cite{BCFII}),
for the calculation of one-loop QCD amplitudes. 
The unitarity-based method applies
most easily to terms in the amplitudes that have discontinuities.
Computation of these terms requires only knowledge of tree amplitudes
evaluated in four dimensions.  The unitarity-based method can also be
applied to computation of rational terms, by computing the cuts at
higher order in $\e$~\cite{UnitarityMachinery,TwoLooppppp}; but doing
so requires knowledge of tree amplitudes with two legs in $D=4-2\e$
dimensions.  These are equivalent to amplitudes with massive
scalars~\cite{UnitarityMachinery,SelfDual}, to which on-shell recursion
relations have also been applied recently~\cite{BadgerMassive}.

This paper is organized as follows.  In \sect{NotationSection}, we
introduce the notation used in the remainder of the paper, and
describe the color organization of one-loop QCD amplitudes with an 
external $q\bar{q}$ pair, in terms of color-stripped objects called
primitive amplitudes.  In \sect{RecursionReviewSection}, we review the
general construction of recursion relations, as well as summarize the known
factorization properties of one-loop amplitudes, which dictate the
structure of the relations.  The recursion relations for the quark 
amplitudes are built using a known set of amplitudes which we present in
\sect{ReviewAmplitudesSection}.  In \sect{AmplitudesSection},
we construct our relations for the finite quark amplitudes,
solve them, and discuss the factorization properties of the solutions.
We provide additional cross checks on the solutions in
\sect{CrossChecksSection}.


\section{Notation}
\label{NotationSection}

We will organize the calculation of one-loop two-quark $(n-2)$-gluon
amplitudes in terms of {\it primitive\/}
amplitudes~\cite{TwoQuarkThreeGluon}.  Each color-ordered amplitude in
a trace-based color
decomposition~\cite{TreeColor,BGSix,MPX,TreeReview,BKColor,TwoQuarkThreeGluon}
is built out of several primitive amplitudes.

For tree-level amplitudes with a quark pair in the fundamental
representation and $(n-2)$ external gluons, the color decomposition
is~\cite{BGSix,MPX,TreeReview},
\begin{equation}
 {\cal A}_n^\tree(1_{\bar{q}},2_q,3,\ldots,n)
 \ =\  g^{n-2} \sum_{\sigma\in S_{n-2}}
   (T^{a_{\sigma(3)}}\ldots T^{a_{\sigma(n)}})_{i_2}^{~\ib_1}\
    A_n^\tree(1_{\bar{q}},2_q;\sigma(3),\ldots,\sigma(n))\,,
\label{TreeColorDecomposition}
\end{equation}
where $S_{n-2}$ is the permutation group on $n-2$ elements,
$j^{h_j}$ denotes the $j$-th momentum $k_j$ and helicity $h_j$,  
and the superscript ``(0)'' signifies that these are leading-order 
tree amplitudes.  The $T^a$
are fundamental representation SU$(N_c)$ color matrices normalized so
that $\Tr(T^a T^b) = \delta^{ab}$.  The color-ordered amplitude
$A_n^\tree$ is related to tree-level all-gluon amplitudes by
supersymmetry Ward identities~\cite{Susy}.  Because the color indices
have been stripped off from the partial amplitudes, there is no need
to distinguish a quark leg $q$ from an anti-quark leg $\bar{q}$;
charge conjugation relates the two choices.

At one loop an additional trace of gluon color matrices $T^{a_i}$ may
survive.  The general color decomposition for fundamental
representation quarks is~\cite{TwoQuarkThreeGluon},
\begin{equation}
 {\cal A}_{n}(1_{\bar{q}},2_q,3,\ldots,n)
 \ =\ { g^n \over (4\pi)^2 } 
  \sum_{j=1}^{n-1} \sum_{\sigma\in S_{n-2}/S_{n;j}}
    \Gr_{n;j}^{(\bar{q}q)}(\sigma(3,\ldots,n))\
  A_{n;j}(1_{\bar{q}},2_q;\sigma(3,\ldots,n))\ ,
\label{OneLoopColorDecomposition}
\end{equation}
where we have extracted a loop factor of $1/(4\pi)^2$, and the color 
structures $\Gr_{n;j}^{(\bar{q}q)}$ are defined by,
\begin{eqnarray}
 \Gr_{n;1}^{(\bar{q}q)}(3,\ldots,n)
  \ &=& N_c\ (T^{a_3}\ldots T^{a_n})_{i_2}^{~\ib_1}\,,\nonumber\\
 \Gr_{n;2}^{(\bar{q}q)}(3;4,\ldots,n)
  \ &=& 0\ ,\label{ColorStructures}\\
 \Gr_{n;j}^{(\bar{q}q)}(3,\ldots,j+1;j+2,\ldots,n)
 \ &=& \Tr(T^{a_3}\ldots T^{a_{j+1}})\ \
   (T^{a_{j+2}}\ldots T^{a_n})_{i_2}^{~\ib_1}\,,
  \quad j=3,\ldots,n-2, \nonumber\\
 \Gr_{n;n-1}^{(\bar{q}q)}(3,\ldots,n)\ &=& \Tr(T^{a_3}\ldots T^{a_n})\ \
     \delta_{i_2}^{~\ib_1} \,. \nonumber
\end{eqnarray}
Here $S_{n;j} = Z_{j-1}$ is the subgroup of $S_{n-2}$ that
leaves $\Gr_{n;j}^{(\bar{q}q)}$ invariant.
When the permutation $\si$ acts on a list of indices, it is to be
applied to each index separately:
$\si(3,\ldots,n) \equiv \si(3),\ldots,\si(n)$, etc.
We refer to $A_{n;1}$ as the leading-color partial amplitude,
and to the $A_{n;j>1}$ as subleading-color,
because for large $N_c$, $A_{n;1}$ alone gives the
leading contribution to the color-summed correction to the cross section,
obtained by interfering ${\cal A}_{n}^\tree$ with ${\cal A}_{n}$.
The explicit $N_c$ in the definition of the leading-color
structure $\Gr_{n;1}^{(\bar{q}q)}$ --- which is otherwise identical to
the tree color structure --- ensures that $A_{n;1}$ is $\Ord(1)$ for
large $N_c$.  (For super-Yang-Mills theory, where the fermions
are adjoint-representation gluinos, one should use the same 
color decomposition as for gluons~\cite{BKColor}. However,
the particular helicity amplitudes considered in this paper 
vanish for super-Yang-Mills theory.)

We describe the amplitudes using the spinor helicity
formalism~\cite{SpinorHelicity,TreeReview}.  In this formalism
amplitudes are expressed in terms of spinor inner-products,
\begin{equation}
\spa{j}.{l} = \langle j^- | l^+ \rangle = \bar{u}_-(k_j) u_+(k_l)\,, 
\hskip 2 cm
\spb{j}.{l} = \langle j^+ | l^- \rangle = \bar{u}_+(k_j) u_-(k_l)\, ,
\label{spinorproddef}
\end{equation}
where $u_\pm(k)$ is a massless Weyl spinor with momentum $k$ and plus
or minus chirality. Our convention
is that all legs are outgoing. The notation used here follows the
standard QCD literature, with $\spb{i}.{j} = \sign(k_i^0 k_j^0)\spa{j}.{i}^*$
so that,
\begin{equation}
\spa{i}.{j} \spb{j}.{i} = 2 k_i \cdot k_j = s_{ij}\,.
\end{equation}
(Note that the square bracket $\spb{i}.{j}$ differs by an overall sign
compared to the notation commonly used in twistor-space
studies~\cite{WittenTopologicalString}.) 

\def\vmu{{\vphantom{\mu}}}
We denote the sums of cyclicly-consecutive external momenta by
\begin{equation}
K^\mu_{i\cdots j} \equiv 
   k_i^\mu + k_{i+1}^\mu + \cdots + k_{j-1}^\mu + k_j^\mu,
\label{KDef}
\end{equation}
where all indices are mod $n$ for an $n$-parton amplitude.
The invariant mass of this vector is $s_{i\cdots j} = K_{i\cdots j}^2$.
Special cases include the two- and three-particle invariant masses, 
which are denoted by
\begin{equation}
s_{ij} \equiv K_{i,j}^2
\equiv (k_i+k_j)^2 = 2k_i\cdot k_j,
\qquad \quad
s_{ijk} \equiv (k_i+k_j+k_k)^2.
\label{TwoThreeMassInvariants}
\end{equation}
In color-ordered amplitudes, only invariants with cyclicly-consecutive
arguments need appear, {\it e.g.}{} $s_{i,i+1}$ and $s_{i,i+1,i+2}$.
We also write, for the sum of massless momenta belonging to a set $A$,
\be
K^\mu_A \equiv \sum_{a_i \in A}
   k_{a_i}^\mu \,.
\label{KDefAlt}
\end{equation}
Spinor strings, such as
\begin{equation}
 \spba{i}.{\Ksl_A}.{j} = \sum_{a\in A} \spb{i}.{a}\spa{a}.j \,, \hskip 2 cm 
 \spab{i}.{\Ksl_A}.{j} = \sum_{a\in A} \spa{i}.{a}\spb{a}.j
  \,,
\label{longerstrings}
\end{equation}
and
\begin{eqnarray}
\spab{i}.{(a+b)}.{j} &=& \spa{i}.{a} \spb{a}.{j} + \spa{i}.{b} \spb{b}.{j} \,,
          \nonumber   \\
\spaa{i}.{(a+b)}.{(c+d)}.{j} &=& 
     \spa{i}.{a} \spba{a}.{(c+d)}.{j} +
     \spa{i}.{b} \spba{b}.{(c+d)}.{j} \,,
\end{eqnarray}
will also make appearances.

%
\begin{figure}[t]
\centerline{\epsfxsize 3.0 truein \epsfbox{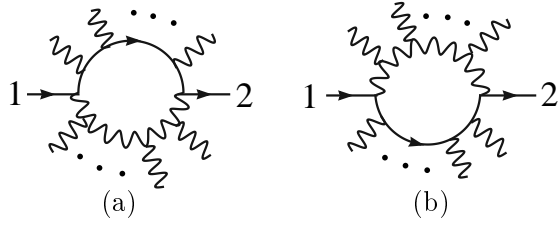}}
\caption{In diagram (a) the fermion line (following the arrow) turns
left on entering the loop and is an $L$ type primitive amplitude.  
In diagram (b) the fermion line turns right and is an $R$ type.}
\label{LeftRightAFigure}
\end{figure}

%
\begin{figure}[t]
\centerline{\epsfxsize 3.0 truein \epsfbox{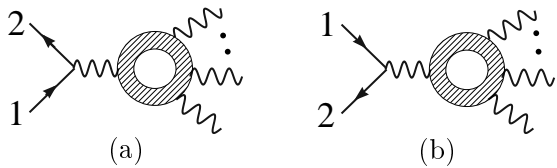}}
\caption{In diagram (a) the external fermion line passes to the ``left''
of the loop, following the fermion arrow, and is designated an $L$
type.  In (b) it passes to the ``right'' and is an $R$ type.  A gluon,
fermion or scalar can circulate in the loop.  The same decomposition
also holds even if we emit additional gluons off the external fermion
lines.}
\label{LeftRightBFigure}
\end{figure}

As noted above, we can write the one-loop color-ordered amplitudes in
the fermionic case more compactly in terms of primitive amplitudes.
Each primitive amplitude corresponds to the set of all color-ordered
diagrams with specified internal states, and a specified orientation
of the fermion line along the loop~\cite{TwoQuarkThreeGluon}.  

First consider the case depicted in \fig{LeftRightAFigure}, where the
fermion line entering the diagram is part of the loop.  Upon entering,
the fermion line can turn either ``left'' and circulate clockwise, or
turn ``right'' and circulate counter-clockwise around the loop.  The
two orientations correspond to separate primitive amplitudes which we
designate as ``$L$'' and ``$R$''.  They would carry the same color factor
were the fermions in the adjoint representation, but carry different
ones when the fermions are in the fundamental representation.  This
division is therefore gauge invariant.  The second case, shown in
\fig{LeftRightBFigure}, is where the external fermion line does not
enter the loop. This case may also be divided into ``left'' and ``right''
pieces, which we again label by ``$L$'' and ``$R$''.  Here the
division is based on whether the loop passes to the left or right of
the fermion line.

The primitive amplitudes for $\qb qgg\cdots g$ are,
\begin{equation}
\matrix{
   & A_n^{L,[J]}(1_\f,3,4,\ldots,j,2_\f,j+1,\ldots,n),  \hfill\cr
   & A_n^{R,[J]}(1_\f,3,4,\ldots,j,2_\f,j+1,\ldots,n),  \hfill\cr}
   \hskip 1 cm  J=1,\coeff{1}{2},0,
\label{PrimitiveAmplitudes}
\end{equation}
where $J={1\over2}$ and $J=0$ denote the contributions with a closed
fermion loop and closed complex scalar loop, respectively.  
The second fermion has been placed in the $j^{\rm th}$ position,
but assigned the label 2.
Since the primitive amplitudes can be used to build amplitudes with any color
representation for the fermions, instead of labelling the fermionic
legs by $q$ and $\bar q$ we label them by $f$ to denote a generic
fermion in any color representation.  The normalization is such that
two helicity states (Weyl fermions or complex scalars) circulate in
the loop.  Diagrams without closed fermion or scalar loops are
assigned to $J=1$; they may or may not contain a closed gluon loop, as
the two types of diagrams mix under gauge transformations.  For
notational simplicity, we shall suppress the superscript ``$[1]$'',
\begin{equation}
A_n^{L} \equiv A_n^{L,[1]}\,, \hskip 2 cm 
A_n^{R} \equiv A_n^{R,[1]}\,.
\end{equation}

The primitive amplitudes~(\ref{PrimitiveAmplitudes}) are not all
independent.  The set of diagrams where the incoming leg $1$ turns left
is related (up to a sign) to a corresponding set where it turns 
right.  This relation is via a reflection which flips over each diagram,
reversing the cyclic ordering:
\begin{equation}
  A_n^{R,[J]}(1_\f,3,4,\ldots,2_\f,\ldots,n-1,n)
   = (-1)^n  A_n^{L,[J]}(1_\f,n,n-1,\ldots,2_\f,\ldots,4,3).
\label{FlipSymmetry}
\end{equation}
In addition, the super-Yang-Mills partial amplitudes for two gluinos and $n-2$
gluons $A_n^\SUSY \equiv A_{n;1}^\SUSY$ are given
by the sum (with all cyclic orderings identical)
\begin{equation}
 A_n^\SUSY\ \equiv\
 A_{n;1}^\SUSY = A_n^{L}\ +\ A_n^{R}\ +\ A_n^{L,[1/2]}\ +\ A_n^{R,[1/2]},
\label{SUSYsum}
\end{equation}
because the ``left'' and ``right''  diagrams have the same group-theory
weight for an adjoint-representation fermion.
Due to supersymmetric cancellations between ``left'' and ``right'' 
primitive amplitudes, $A_n^\SUSY$ is always simpler
than either $A_n^{L}$ or $A_n^{R}$.
\Eqn{SUSYsum} allows one to obtain one of the four terms on the right
with no effort, given $A_n^\SUSY$.  
In the case of the finite amplitudes we are considering
in this paper, $A_n^\SUSY$ vanishes.  
We will choose to eliminate $A_n^R$, and compute $A_n^L$.

Finally, the following fermion-loop contributions vanish,
\begin{eqnarray}
 A_n^{R,[1/2]}(1_\f,2_\f,3,4,\ldots,n) &=&
 A_n^{L,[1/2]}(1_\f,n,\ldots,4,3,2_\f)  = 0, \nonumber\\
 A_n^{R,[1/2]}(1_\f,3,2_\f,4,\ldots,n) &=&
 A_n^{L,[1/2]}(1_\f,n,\ldots,4,2_\f,3) = 0,  \label{FLoopVanish}
\end{eqnarray}
and similarly for the scalar-loop contributions.  The restriction to
``left'' or ``right'' diagrams combines with the ordering of the
external legs to leave only tadpole and massless external bubble
diagrams behind; but these are zero in dimensional regularization.

For scalars and fermions in the fundamental representation circulating
in the loop, the leading-color contribution to 
\eqn{OneLoopColorDecomposition},
$A_{n;1}$, is given in terms of primitive amplitudes by, 
\begin{eqnarray}
A_{n;1}(1_{\bar{q}},2_q;3,\ldots,n) &= &
   A_n^L(1_\f,2_\f,3,\ldots,n)
  - {1\over N_c^2} A_n^R(1_\f,2_\f,3,\ldots,n) \nonumber\\
 &&  + {\nf\over N_c} A_n^{L,[1/2]}(1_\f,2_\f,3,\ldots,n)
  + {\ns\over N_c} A_n^{L,[0]}(1_\f,2_\f,3,\ldots,n) \,. 
\label{Anoneformula}
\end{eqnarray}
For QCD the number of scalars vanishes, $\ns = 0$, while $\nf$ is the number of
light quark flavors. (Our fundamental representation scalars are normalized
so that they carry the same number of states as Dirac fermions, $4 N_c$.) 

As explained in ref.~\cite{TwoQuarkThreeGluon}, the subleading-color
partial amplitudes $A_{n;j>1}$ appearing in \eqn{OneLoopColorDecomposition}
may be expressed as a permutation sum over primitive amplitudes,
\begin{eqnarray}
&& \hskip -3 cm A_{n;j}(1_\qb,2_q; 3,\ldots,j+1;j+2,j+3,\ldots,n)  \nonumber\\
\hskip 2 cm   & = &
 (-1)^{j-1} \sum_{\sigma\in COP\{\alpha\}\{\beta\}}
    \Biggl[ A_n^{L,[1]} ( \sigma(1_\f,2_\f,3,\ldots,n) ) \nonumber \\
 &&   - {\nf\over N_c} A_n^{R,[1/2]} (\sigma(1_\f,2_\f,3,\ldots,n) )
   - {\ns\over N_c} A_n^{R,[0]} (\sigma(1_\f,2_\f,3,\ldots,n) )
     \Biggr]\ , 
\label{subltotal}
\end{eqnarray}
where $\alpha_i \in \{\alpha\} \equiv \{j+1,j,\ldots,4,3\}$,
$\beta_i \in \{\beta\} \equiv \{1,2,j+2,j+3,\ldots,n-1,n\}$,
and  $COP\{\alpha\}\{\beta\}$ is the set of all
permutations of $\{1,2,\ldots,n\}$ with leg $1$ held fixed
that preserve the cyclic
ordering of the $\alpha_i$ within $\{\alpha\}$ and of the $\beta_i$
within $\{\beta\}$, while allowing for all possible relative orderings
of the $\alpha_i$ with respect to the $\beta_i$.
For example if $\{\alpha\} = \{4,3\}$ and
$\{\beta\} = \{1,2,5\}$ (the case required for $A_{5;3}$), 
then $COP\{\alpha\}\{\beta\}$ contains the twelve elements
\begin{eqnarray}
& & (1,2,5,4,3),\quad (1,2,4,5,3),\quad (1,4,2,5,3),\quad
    (1,2,4,3,5),\quad \nonumber\\
&&  (1,4,3,2,5),\quad (1,4,2,3,5),\quad
    (1,2,5,3,4),\quad (1,2,3,5,4),\quad  \label{OrderingExample}\\
&&  (1,3,2,5,4),\quad (1,2,3,4,5),\quad (1,3,4,2,5),\quad (1,3,2,4,5). \quad
\nonumber
\end{eqnarray}
Thus, all partial amplitudes appearing in
\eqn{OneLoopColorDecomposition} are expressible as sums over primitive
amplitudes and it is sufficient to compute the primitive amplitudes in
order to fully specify the complete color-dressed amplitudes.

For the case at hand, where all external gluons carry
the same helicity, a supersymmetry identity~\cite{Susy} may be used to
prove that the fermion loop and scalar loop are the same up to a sign,
\begin{eqnarray}
A_n^{L,[0]}(1_\f^-,2^+,\ldots, j_\f^+, \ldots,n^+) 
& = &- A_n^{L,[1/2]}(1_\f^-,2^+,\ldots, j_\f^+, \ldots,n^+) 
  \nonumber \\
& \equiv & A_n^{s}(1_\f^-,2^+,\ldots, j_\f^+, \ldots,n^+) \,.
\label{AsDef}
\end{eqnarray}
Thus, by computing the closed scalar-loop primitive amplitude
we obtain also the closed fermion-loop primitive amplitude.

We shall find it convenient to compute the combinations
\begin{equation}
A_n^s\,, \hskip 1cm \hbox{and} \hskip 1cm  A_n^{L-s} \equiv A_n^L-A_n^s\,,
\label{LminusSDef}
\end{equation}
instead of $A_n^s$ and $A_n^L$.

Furthermore, combining reflection symmetry and supersymmetry leads to,
\begin{eqnarray}
&&\hskip - 1.7 cm  A_n^L(1_\f^-,2^+,\ldots,j_\f^+,\ldots,n^+) =
(-1)^n A_n^R(1_\f^-,n^+,\ldots,j_\f^+,\ldots,2^+)\nonumber\\
\hskip 1 cm \null &=& (-1)^n \Bigl[
A_n^{\rm SUSY}(1_\f^-,n^+,\ldots,j_\f^+,\ldots,2^+)
-A_n^L(1_\f^-,n^+,\ldots,j_\f^+,\ldots,2^+)
\nonumber\\ &&\hphantom{= (-1)^n \Bigl[]} 
  -A_n^{L,[\onehalf]}(1_\f^-,n^+,\ldots,j_\f^+,\ldots,2^+)
  -A_n^{R,[\onehalf]}(1_\f^-,n^+,\ldots,j_\f^+,\ldots,2^+)\Bigr]
\nonumber\\
&=& (-1)^n \Bigl[
-A_n^L(1_\f^-,n^+,\ldots,j_\f^+,\ldots,2^+)
\\ &&\hphantom{= (-1)^n \Bigl[]} 
  +A_n^{L,[0]}(1_\f^-,n^+,\ldots,j_\f^+,\ldots,2^+)
  +A_n^{R,[0]}(1_\f^-,n^+,\ldots,j_\f^+,\ldots,2^+)\Bigr]
\nonumber\\
&=& (-1)^n \Bigl[
-A_n^L(1_\f^-,n^+,\ldots,j_\f^+,\ldots,2^+)
\nonumber\\ &&\hphantom{= (-1)^n \Bigl[]} 
  +A_n^{s}(1_\f^-,n^+,\ldots,j_\f^+,\ldots,2^+)
  + (-1)^n A_n^{s}(1_\f^-,2^+,\ldots,j_\f^+,\ldots,n^+)\Bigr],
\nonumber
\end{eqnarray}
so that
\begin{equation}
A_n^{L-s}(1_\f^-,2^+,\ldots,j_\f^+,\ldots,n^+) =
(-1)^{n+1} A_n^{L-s}(1_\f^-,n^+,\ldots,j_\f^+,\ldots,2^+),
\label{Lminussflip}
\end{equation}
which we can use to obtain $A_n^{L-s}$ for $j>\lceil (n+1)/2\rceil$
($\lceil x\rceil$ is the smallest integer greater than or equal to $x$).
Note also that, as discussed above,
$A_n^s$ vanishes if $j=n$ or $j=n-1$.

In summary, for amplitudes with a single quark pair
with identical helicity gluon legs, there are two independent classes
of primitive amplitudes that need to be computed,
\begin{equation}
A_n^{L-s}(1_\f^-,2^+,\ldots, j_\f^+, \ldots,n^+) \,,
\hskip 1cm \hbox{and} \hskip 1 cm 
A_n^{s}(1_\f^-,2^+,\ldots, j_\f^+, \ldots,n^+) \,.
\end{equation}
%


\section{Review of Recursion Relations}
\label{RecursionReviewSection}

\subsection{On-Shell Recursion Relations for Trees}

The on-shell recursion relations rely on general properties
of complex functions as well as factorization properties of 
scattering amplitudes. 
The proof~\cite{BCFW} of the relations relies on a
parameter-dependent shift of two of the external massless spinors,
here labelled $k$ and $l$,
in an $n$-point process,
\begin{eqnarray}
&\tlambda_k &\rightarrow \tlambda_k - z\tlambda_l \,, \nonumber\\
&\lambda_l &\rightarrow \lambda_l + z\lambda_k \,,
\label{SpinorShift}
\end{eqnarray}
where $z$ is a complex number.  The corresponding momenta
(labeled by $p_i$ instead of $k_i$ in this section) 
are shifted as well,
\begin{eqnarray}
&p_k^\mu &\rightarrow p_k^\mu(z) = p_k^\mu - 
      {z\over2}{\sand{k}.{\gamma^\mu}.{l}},\nonumber\\
&p_l^\mu &\rightarrow p_l^\mu(z) = p_l^\mu + 
      {z\over2}{\sand{k}.{\gamma^\mu}.{l}} \,,
\label{MomentumShift}
\end{eqnarray}
so that they remain massless, $p_k^2(z) = 0 = p_l^2(z)$,
and overall momentum conservation is maintained.
The shift also implies,
\begin{eqnarray}
&\psl_k &\rightarrow 
  \psl_k(z) = \psl_k 
  - z \, 
  \bigl( \, {|l^-\rangle\langle k^-| + |k^+\rangle\langle l^+|} \, \bigr)\,,
\nonumber\\
&\psl_l &\rightarrow 
  \psl_l(z) = \psl_l
  + z \, 
  \bigl( \, {|l^-\rangle\langle k^-| + |k^+\rangle\langle l^+|} \, \bigr) \,.
\label{SlashedMomentumShift}
\end{eqnarray}

Define a parameter-dependent continuation of an on-shell amplitude,
\begin{equation}
A(z) = A(p_1,\ldots,p_k(z),p_{k+1},\ldots,p_l(z),\ldots,p_n),
\end{equation}
evaluated at a particular set of complex
momenta.  When $A$ is a tree amplitude or finite one-loop
amplitude, $A(z)$ is a rational function of $z$.
The physical amplitude is given by $A(0)$. 

Consider the contour integral, 
\begin{equation}
{1\over 2\pi i} \oint_C {dz\over z}\,A(z) \,,
\label{ContourInt}
\end{equation}
where the contour is taken around the circle at infinity.  If
$A(z)\rightarrow 0$ as $z\rightarrow\infty$, as in the tree-level
cases~\cite{BCFRecurrence,BCFW,LuoWen,GravityRecurrence,BadgerMassive},
then there is no ``surface term''; that is, the
integral~(\ref{ContourInt}) vanishes. Evaluating the integral as a sum
of residues, we can then solve for $A(0)$ to obtain,
\begin{equation}
A(0) = -\sum_{{\rm poles}\ \alpha} \Res_{z=z_\alpha}  {A(z)\over z}\,.
\label{NoSurface}
\end{equation}

As explained in ref.~\cite{BCFW}, if $A(z)$ only has simple poles,
each residue is given by factorizing the shifted amplitude on the
appropriate pole in momentum invariants, so that at tree level,
\begin{equation}
A(0) = \sum_{r,s,h} 
   A^h_L(z = z_{rs}) { i \over K_{r\cdots s}^2 } A^{-h}_R(z = z_{rs})  \,,
\label{BCFWRepresentation}
\end{equation}
where $h=\pm1$ labels the helicity of the intermediate state.  There
is generically a double sum, labeled by $r,s$, over momentum poles,
with legs $k$ and $l$ always appearing on opposite sides of the pole.
The squared momentum associated with that pole, $K_{r\cdots s}^2$, is
evaluated in the unshifted kinematics; whereas the on-shell amplitudes
$A_L$ and $A_R$ are evaluated in kinematics that have been shifted by
\eqn{SpinorShift} with $z=z_{rs}$, where
\begin{equation}
z_{rs} = - {K_{r\cdots s}^2 \over \sand{k}.{\Ksl_{r\ldots s}}.{l} } \,.
\end{equation}
To extend the approach to one loop~\cite{OnShellRecurrenceI}, 
the sum (\ref{BCFWRepresentation})
should also be taken over the two ways of assigning the loop to $A_L$
and $A_R$.  This formula assumes that there are no additional poles
present in the amplitude other than the standard poles for real
momenta.  At tree level it is possible to demonstrate the absence of
additional poles, but at loop level it is not true.

For the case of a fermionic pole, there is a sign subtlety similar to
the situation with maximally-helicity-violating (MHV)
vertices~\cite{Currents}.  The correct
fermionic propagator is, of course, $i \ksl/k^2$.  As is the case for
MHV vertices, the $\ksl$ is supplied by the amplitudes appearing on
both sides of the pole.  In these amplitudes, each momentum is
directed outwards.  Thus when we link two amplitudes across the pole
we would obtain a numerator factor of the form $|k_1^+\rangle \langle
k_2^+ |$, where $k_2 = -k_1$.  This is not quite right since the same
momentum argument should appear in both spinors, {\it i.e.,} $|k_1^+
\rangle \langle k_1^+ | \neq |k_1^+ \rangle \langle (-k_1)^+|$.  To
correct this we flip the sign of the momentum in the
spinor $\langle k_2^+ |$.

Because of the general structure of multiparticle
factorization~\cite{BernChalmers}, only standard single poles in $z$
arise from multiparticle channels, even at one loop.  However, as was
pointed out in ref.~\cite{OnShellRecurrenceI}, double poles in $z$ do
arise at one loop due to collinear factorization.  The splitting
amplitudes with helicity configuration $(+{}+{}+)$ and $(-{}-{}-)$ (in 
an all-outgoing helicity convention) can
lead to double poles in $z$, because their dependence on the spinor
products takes the form $\spb{a}.{b}/\spa{a}.{b}^2$ for $(+{}+{}+)$,
or its complex conjugate $\spa{a}.{b}/\spb{a}.{b}^2$ for
$(-{}-{}-)$~\cite{Neq4Oneloop}.  As discussed in
ref.~\cite{OnShellRecurrenceI}, this behavior alters the form of the
recursion relation in an essential manner.  In general, underneath
the double pole sits an object of the form,
\begin{equation}
{\spb{a}.b\over\spa{a}.b}\,,
\label{UnrealPole}
\end{equation}
which we call an ``unreal pole'' since there is no pole present when
real momenta are used; it only appears, as a single pole, when we
continue to complex momenta.  As we shall discuss in
\sect{AmplitudesSection}, the finite quark amplitudes exhibit similar
phenomena, except that in the quark case we shall also encounter
unreal poles not directly associated with a double pole.

\subsection{One-loop Factorization Properties}
\label{OneLoopFact}

In order to build on-shell recursion relations, we need the
factorization properties of one-loop amplitudes for complex momenta.
It is useful to first review the factorization properties for real
momenta, which we know from general
arguments~\cite{TreeReview,BernChalmers,OneloopSplit}.  Although a
good starting point, this is in general insufficient due to the
appearance of unreal poles.

As the real momenta of two external legs become collinear, any
one-loop amplitude in massless gauge theory will factorize as, 
\begin{eqnarray}
 A_{n}^{\oneloop} \inlimit^{a \parallel b}\
\sum_{\lambda=\pm}  && \null \hskip -.35 cm \biggl(
\Split^\tree_{-\lambda}   (a^{\lambda_a},b^{\lambda_b};z)\,
         A_{n-1}^{\oneloop}(\ldots,(a+b)^\lambda,\ldots) \nonumber\\
&&\hphantom{\biggl()}
 + \Split^{\oneloop}_{-\lambda}(a^{\lambda_a},b^{\lambda_b};z)\,
         A_{n-1}^\tree(\ldots,(a+b)^\lambda,\ldots) \biggr)\,,
\end{eqnarray}
where $a$ and $b$ are nearest neighbors in the cyclic ordering of
legs. (When they are not nearest neighbors, there is no
universal factorization behavior, but also no collinear singularity.) 
There are three distinct kinds of
collinear limits to consider: two gluons becoming collinear; a gluon
becoming collinear with a quark (or anti-quark); and a quark and
anti-quark becoming collinear (in the case that they are adjacent).
These limits simplify in the case at hand, where all external gluons
have positive helicity, due to vanishing of relevant $(n-1)$-point
tree amplitudes.  The required splitting amplitudes are tabulated in
ref.~\cite{Neq4Oneloop}.

First consider the case when two cyclically adjacent gluons, $a$ and $b$, 
become collinear. In this case $A_n^L$ behaves as,
\begin{eqnarray}
 A_{n}^{L}(\ldots,a^+,b^+,\ldots) \inlimit^{a \parallel b}\
&&
\Split^\tree_{-}   (a^{+},b^{+};z)\,
         A_{n-1}^{L}(\ldots,(a+b)^+,\ldots) \nonumber\\
&& + \Split^{\oneloop:g}_{+}(a^{+},b^{+};z)\,
         A_{n-1}^\tree(\ldots,(a+b)^-,\ldots) \,,
\label{TwoGluonLimitL}
\end{eqnarray}
where $\Split^{\oneloop:g}$ is the one-loop splitting amplitude
with a gluon circulating in the loop.  Similarly,
$A_n^s$ behaves as, 
\begin{eqnarray}
 A_{n}^{s}(\ldots,a^+,b^+,\ldots) \inlimit^{a \parallel b}\
&&
\Split^\tree_{-}   (a^{+},b^{+};z)\,
         A_{n-1}^{s}(\ldots,(a+b)^+,\ldots) \nonumber\\
&& + \Split^{\oneloop:s}_{+}(a^{+},b^{+};z)\,
         A_{n-1}^\tree(\ldots,(a+b)^-,\ldots) \,,
\label{TwoGluonLimits}
\end{eqnarray}
where $\Split^{\oneloop:s}$ is the one-loop splitting amplitude with a
scalar circulating in the loop.  In both cases, the remaining two
terms, with opposite intermediate-gluon helicity, vanish because
$\Split^\tree_{+}(a^{+},b^{+};z)$ and
$A_{n-1}^\tree(\pm,+,+,\ldots,+)$ are zero.  Now~\cite{Neq4Oneloop},
\begin{equation}
\Split^{\oneloop:g}_{+}(a^{+},b^{+};z) = 
\Split^{\oneloop:s}_{+}(a^{+},b^{+};z),
\end{equation}
thanks to a supersymmetry Ward identity~\cite{Susy}.  
Taking the difference we see that in this limit the  $A_{n}^{L-s}$ class
of amplitudes to be computed has a very simple structure,
\begin{equation}
 A_{n}^{L-s}(\ldots,a^+,b^+,\ldots) \inlimit^{a \parallel b}\
\Split^\tree_{-}   (a^{+},b^{+};z)\,
         A_{n-1}^{L-s}(\ldots,(a+b)^+,\ldots) \,.
\label{DeltaLimitI}
\end{equation}

In the limit that a gluon becomes collinear with the negative-helicity
fermion, for either $A_n^L$ or $A_n^s$ we find, thanks to helicity
conservation on a fermion line, that,
\begin{equation}
 A_{n}^{L,s}(a_\f^-,b^+,\ldots) \inlimit^{a \parallel b}
\Split^\tree_{(f)+}(a_\f^{-},b^{+};z)\,
         A_{n-1}^{L,s}((a+b)_\f^-,\ldots) \,.
\end{equation}
Again taking the difference, we obtain,
\begin{equation}
 A_{n}^{L-s}(a_\f^-,b^+,\ldots) \inlimit^{a \parallel b} 
\Split^\tree_{(f)+}(a_\f^{-},b^{+};z)\,
         A_{n-1}^{L-s}((a+b)_\f^-,\ldots) \,.
\label{DeltaLimitII}
\end{equation}
A similar expression holds for the limit in which a gluon becomes collinear 
with the positive-helicity fermion.

Finally, if the quark and anti-quark are adjacent, the amplitude
factorizes onto the finite one-loop pure-glue amplitudes,
\begin{eqnarray}
 A_{n}^{L,s}(a_\f^-,b_\f^+,\ldots) \inlimit^{a \parallel b}\
&&
\Split^\tree_{-}(a_\f^-,b_\f^+;z)\,
      A_{n-1}^{\oneloop}((a+b)^+,\ldots) \nonumber\\
&& \null  + \Split^\tree_{+}(a_\f^-,b_\f^+;z)\,
         A_{n-1}^\oneloop((a+b)^-,\ldots)\,.
\label{QuarkAntiQuarkLimit}
\end{eqnarray}
The gluon and scalar loop contributions to $A_{n-1}^\oneloop(\pm,+,+,\ldots,+)$
are the same by supersymmetry~\cite{Susy}.  Hence the 
difference $A_n^{L-s}$ is finite in this limit,
\begin{equation}
 A_{n}^{L-s}(a_\f^-,b_\f^+,\ldots) \inlimit^{a \parallel b}\
\hbox{non-singular}\,.
\label{DeltaLimitIII}
\end{equation}

In the case of multiparticle factorization, the vanishing of
$A_n^\tree(1_\qb^-,2_q^+,3^+,\ldots,n^+)$ implies that
we can only factorize on a gluon pole,
\begin{eqnarray}
&& \null \hskip -4 mm 
A_n^{L,s}(1_\f^-,2^+,\ldots,j_\f^+,\ldots,m^+,\ldots,n^+) 
\inlimit^{K_{1\cdots m}^2\rightarrow 0}
\\ &&\hskip 5mm
A_{m+1}^\tree(1_\f^-,2^+,\ldots,j_\f^+,\ldots,m^+,(-K_{1\ldots m})^-)
{i \over K_{1\ldots m}^2 }
A_{n-m+1}^{\oneloop}(K_{1\ldots m}^+,(m+1)^+,\ldots,n^+) \,,
\nonumber
\end{eqnarray}
with $m\ge3$.  Again $A_n^{L-s}$ has a non-singular limit,
\begin{equation}
A_n^{L-s}(1_\f^-,2^+,\ldots,j_\f^+,\ldots,m^+,\ldots,n^+) 
\inlimit^{K_{1\cdots m}^2\rightarrow 0}
  {\rm non\dash{}singular} \,,
\label{DeltaLimitIV}
\end{equation}
since once again the behavior of the $L$ and $s$ pieces are identical
and cancel in the $L-s$ difference.

Because of the appearance of unreal poles (\ref{UnrealPole}) this is
not the entire story.  Unfortunately, as yet there are no general
theorems to guide us on the factorization properties in this new class
of poles.  As shown in ref.~\cite{OnShellRecurrenceI} for the finite
one-loop pure-glue amplitudes, factorization on the unreal poles 
is not solely into products of lower-point amplitudes; 
other factors arise.  For the finite quark amplitudes we shall find, 
just as in ref.~\cite{OnShellRecurrenceI}, a systematic set of correction
factors.  In \sect{AmplitudesSection} we will comment on different 
types of factors that appear.

\section{Review of Known Finite QCD amplitudes}
\label{ReviewAmplitudesSection}

In this section, we collect previously-known results for tree and one-loop
finite amplitudes, which feed into the recursive formul\ae{} for the
finite quark amplitudes to be discussed in~\sect{AmplitudesSection}.

We will need the tree-level MHV amplitudes~\cite{ParkeTaylor,TreeReview}, 
\begin{equation}
A_n^\tree(1^+, 2^+, \ldots, m_1^-, \ldots, m_2^-, \ldots, n^+) = 
     i { \spa{m_1}.{m_2}^4 \over \spa1.2 \spa2.3 \cdots \spa{n}.1} \,,
\label{mhvtree}
\end{equation}
and
\begin{equation}
A_n^\tree(1_\f^-, 2^+, \ldots, j_\f^+, \ldots,  m^-, \ldots, n^+) = 
     i { \spa{1}.{m}^3 \spa{j}.{m} \over \spa1.2 \spa2.3 \cdots \spa{n}.1} \,,
\label{ffmhvtree}
\end{equation}
where we use the generic fermion label $f$ again, and
where the omitted labels refer to positive-helicity gluons.

We will also need the one-loop pure-gluon amplitudes, either with 
all helicities positive, or with a single negative helicity. 
For the all-positive case with $n\ge4$ legs, 
Chalmers and the authors~\cite{AllPlusA,AllPlus} wrote a 
conjecture~\footnote{%
Note that a version of the ``odd'' terms $O_n$ in the first reference
in ref.~\cite{AllPlus} (the last line of eq.~(7)) has the wrong sign.}
based on collinear limits,
\begin{equation}
A_n^\oneloop(1^+, 2^+, \ldots, n^+) = {i \over 3} 
{H_n\over \spa1.2 \spa2.3\cdots \spa{(n-1)}.n\spa{n}.1} \,,
\label{OneLoopAllPlusAmplitude}
\end{equation}
where
\begin{equation}
H_n = -\sum_{1\leq i_1<i_2<i_3<i_4\leq n} 
  \Tr_{-}\Bigl[\ksl_{i_1}\ksl_{i_2}\ksl_{i_3}\ksl_{i_4}\Bigr] \,,
\label{Hndef}
\end{equation}
and 
\begin{eqnarray}
\Tr_-\Bigl[\ksl_{i_1}\ksl_{i_2}\ksl_{i_3}\ksl_{i_4}\Bigr] &=& {1\over 2}
\Tr[(1-\gamma_5)\ksl_{i_1}\ksl_{i_2}\ksl_{i_3}\ksl_{i_4}] \nonumber\\
&=& \spa{i_1}.{i_2} \spb{i_2}.{i_3} \spa{i_3}.{i_4} \spb{i_4}.{i_1} \,.
\end{eqnarray}
The conjecture was proven by Mahlon~\cite{Mahlon}.  In
\eqn{OneLoopAllPlusAmplitude} we have extracted an overall factor of
$1/(4\pi)^2$ compared with ref.~\cite{AllPlus}, 
to be consistent with the normalization in
\eqn{OneLoopColorDecomposition} for the quark amplitudes, and we have
set a multiplicity-counting parameter $N_p$ to 2, to match the number
of color-stripped bosonic states defined to be circulating in the loop
in the amplitudes $A_n^s$. More generally, the overall prefactor $N_p$
is just the difference between the number of bosonic and fermionic states
circulating in the loop.

For $n=3$ the proper vertex where legs $1$ and $2$ are the external legs
is~\cite{OnShellRecurrenceI},
\begin{eqnarray}
A_3^\oneloop(1^+, 2^+, 3^+) & = &
 - {i \over 3}
  {\spb1.2 \spb2.3 \spb3.1  \over K_{12}^2} \,.
\label{OneLoopThreeAmplitude}
\end{eqnarray}
It is useful to expose the kinematic pole, so we define the vertex,
\begin{equation}
V_3^\oneloop(1^+, 2^+, 3^+)  \equiv K_{12}^2\,
              A_{3}^\oneloop(1^+, 2^+, 3^+)
= - {i \over 3} \spb1.2 \spb2.3 \spb3.1 \,.
\label{OneLoopThreeVertex}
\end{equation}

The single negative-helicity amplitudes may also 
be required as inputs into the quark recursion relation. 
The four-point amplitude was first calculated using string-based methods and 
is given by~\cite{BKStringBased},
\begin{eqnarray}
A_{4}^\oneloop (1^-,2^+,3^+,4^+)  &=& 
 {i \over 3} \,
     {\spa2.4 \spb2.4^3 \over \spb1.2 \spa2.3 \spa3.4 \spb4.1}\,.
\label{mppp}
\end{eqnarray}
The five-point amplitude with a single negative-helicity leg was also
first calculated using string-based methods and is given
by~\cite{GGGGG,AllPlusA},
\begin{equation}
A_{5}^\oneloop (1^-, 2^+, 3^+, 4^+, 5^+) = 
 {i \over 3} \,  {1\over \spa3.4^2} 
\Biggl[-{\spb2.5^3 \over \spb1.2 \spb5.1}
       + {\spa1.4^3 \spb4.5 \spa3.5 \over \spa1.2 \spa2.3 \spa4.5^2}
       - {\spa1.3^3 \spb3.2 \spa4.2 \over \spa1.5 \spa5.4 \spa3.2^2} 
     \Biggr] \,. 
\label{mppppsimple}
\end{equation}
The $n$-point generalization was then obtained by Mahlon~\cite{Mahlon}
via off-shell recursive methods.  We recently used an on-shell
recursion relation to obtain compact representations of the six- and
seven-point amplitudes~\cite{OnShellRecurrenceI}. The six-point amplitude is,
\begin{eqnarray}
&&A_{6}^\oneloop(1^-,2^+,3^+,4^+,5^+,6^+)\nonumber \\ 
&& \hskip0.5cm =
 {i \over 3} \, \Biggl[
 { {\spab1.{(2+3)}.6}^3
    \over \spa1.2 \spa2.3 {\spa4.5}^2 \, s_{123} \, \spab3.{(1+2)}.6 }
\null + { {\spab1.{(3+4)}.2}^3
    \over {\spa3.4}^2 \spa5.6 \spa6.1 \, s_{234} \, \spab5.{(3+4)}.2 }
\nonumber \\ 
&& \hskip2.0cm
\null + { {\spb2.6}^3 \over \spb1.2 \spb6.1 \, s_{345} } \Biggl(
      { \spb2.3 \spb3.4 \over \spa4.5 \, \spab5.{(3+4)}.2 }
    - { \spb4.5 \spb5.6 \over \spa3.4 \, \spab3.{(1+2)}.6 }
    + { \spb3.5 \over \spa3.4 \spa4.5 } \Biggr)
\nonumber \\ 
&& \hskip2.0cm
\null - { {\spa1.3}^3 \spb2.3 \spa2.4
     \over {\spa2.3}^2 {\spa3.4}^2 \spa4.5 \spa5.6 \spa6.1 }
+ { {\spa1.5}^3 \spa4.6 \spb5.6
     \over \spa1.2 \spa2.3 \spa3.4 {\spa4.5}^2 {\spa5.6}^2 }
\nonumber \\ 
&& \hskip2.0cm
- { {\spa1.4}^3 \spa3.5 \spab1.{(2+3)}.4
     \over \spa1.2 \spa2.3 {\spa3.4}^2 {\spa4.5}^2 \spa5.6 \spa6.1 } 
\Biggr]  \,.
\label{mpppppsimple}
\end{eqnarray}
In \sect{AmplitudesSection}, we give a compact expression for all $n$.

The known results for the fermionic four-point
amplitudes~\cite{KunsztEtAl} are,
\begin{eqnarray}
A_4^{L-s}(1_\f^-,2_\f^+,3^+,4^+) &=&
         -{i\over2} {\spa1.2\spb2.4\over\spa2.3\spa3.4} \,,\nonumber\\
A_4^s(1_\f^-,2_\f^+,3^+,4^+) &=&
  -{i\over3} {\spa1.4\spb2.4\over \spa3.4^2}\,,
\label{FourPoint}
\end{eqnarray}
and for the five-point fermionic amplitudes~\cite{TwoQuarkThreeGluon},
\begin{eqnarray}
A_5^{L-s}(1_\f^-,2_\f^+,3^+,4^+,5^+) &=& 
{i\over2} {\spa1.2\spb2.3\spa3.1+\spa1.4\spb4.5\spa5.1\over
           \spa2.3\spa3.4\spa4.5\spa5.1} \,,
\nonumber\\
A_5^s(1_\f^-,2_\f^+,3^+,4^+,5^+) &=&
 -{i\over3} \Biggl( {\spa1.3\spb3.4\spa4.1^2\over\spa1.2\spa3.4^2\spa4.5\spa5.1}
       +{\spa1.4\spa2.4\spb4.5\spa5.1\over\spa1.2\spa2.3\spa3.4\spa4.5^2}
\nonumber\\
&&\hskip10mm
       +{\spb2.3\spb2.5\over\spb1.2\spa3.4\spa4.5}
            \Biggr)\,,\nonumber\\
A_5^{L-s}(1_\f^-,2^+,3_\f^+,4^+,5^+) &=&
{i\over2} {\spa1.3(\spa1.2\spb2.3\spa3.1+\spa1.4\spb4.5\spa5.1)
           \over \spa1.2\spa2.3\spa3.4\spa4.5\spa5.1} \,,\nonumber\\
A_5^s(1_\f^-,2^+,3_\f^+,4^+,5^+) &=&
{i\over3} {\spa1.4\spa1.5\spb4.5\over\spa1.2\spa2.3\spa4.5^2}\,.
\label{FivePoint}
\end{eqnarray}

\section{Quark Recursion Relations}
\label{AmplitudesSection}

As was discussed in \sect{NotationSection}, for the finite
helicity amplitudes --- one quark pair, with all gluons having
positive helicity --- there are two independent primitive amplitudes
that need to be computed,
\begin{eqnarray}
A_n^{L-s}(j_\f^+) &\equiv& A_n^{L-s}(1_\f^-,2^+,\ldots,j_\f^+,\ldots,n^+) \,,
\qquad 2\leq j\leq \lceil (n+1)/2\rceil\,,
\label{Lminussshort} \\
A_n^{s}(j_\f^+) &\equiv& A_n^{s}(1_\f^-,2^+,\ldots,j_\f^+,\ldots,n^+) \,,
\qquad 2\leq j \leq n-2\,,
\label{sshort}
\end{eqnarray}
where we have adopted an abbreviated notation retaining only the label
of the positive-helicity fermion.  A computation of these primitive
amplitudes then determines all the finite one-loop quark amplitudes in
QCD.

\subsection{Structure of Five-Point Recursion for $L-s$ Contribution}

Let us begin, as in ref.~\cite{OnShellRecurrenceI}, by examining
the structure of five point amplitudes.  Consider
$A_5^{L-s}(2_\f^+)$, as defined in \eqn{Lminussshort}, and
choosing $(k,l)=(1,5)$ as shift variables in \eqn{SpinorShift},
\begin{eqnarray}
\lambda_1 &\rightarrow& \lambda_1 \,,\nonumber\\
\tlambda_1 &\rightarrow& \tlambda_1 - z\tlambda_5 \,,\nonumber\\
\lambda_5 &\rightarrow& \lambda_5 + z\lambda_1 \,,\\
\tlambda_5 &\rightarrow& \tlambda_5\,,\nonumber
\end{eqnarray}
with all other spinors unchanged.  Under this shift, using the known
result for the amplitude we have
\begin{equation}
A_5^{L-s}(2_\f^+; z) = 
{i \over 2}
{\spa1.2\spa1.3\spb2.3\over\spa2.3\spa3.4(\spa4.5 + z \spa4.1) \spa1.5}
+ {i \over 2}
{\spa1.4\spb4.5\over\spa2.3\spa3.4 (\spa4.5 + z  \spa4.1)} \,.
\end{equation}
which vanishes as $z \rightarrow \infty$, so no surface term is required.
Looking at the $z=0$ pole in $A_5^{L-s}(z)/z$, 
we expect to find the following two terms,
\begin{equation}
{i \over 2}
{\spa1.2\spa1.3\spb2.3\over\spa2.3\spa3.4\spa4.5\spa1.5}
+ {i \over 2}
{\spa1.4\spb4.5\over\spa2.3\spa3.4\spa4.5} \,,
\label{A5decomposition}
\end{equation}
appearing in the recursion.

%
\begin{figure}[t]
\centerline{\epsfxsize 1.8 truein \epsfbox{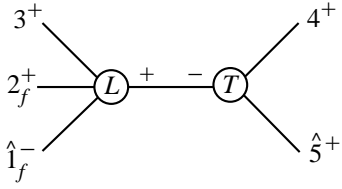}}
\caption{The real-pole diagram in the recursion relation for 
$A_5^{L-s}(1_\f^+, 2_\f^+,3^+,4^+,5^+)$.  The vertices labeled by a $T$ 
are trees, and the ones labeled by an $L$ are loops. }
\label{A5Lms15Figure}
\end{figure}

On the other hand, from the structure of the collinear 
limits --- eqs.~(\ref{DeltaLimitI}), (\ref{DeltaLimitII}), 
and (\ref{DeltaLimitIII}) --- we expect to find only one term.  
A lone term, displayed in \fig{A5Lms15Figure}, is indeed what emerges
from the naive form of the recursion relation,
\begin{eqnarray}
D_1 &=&{i\over s_{45}} A_4^{L-s}(\hat 1_\f^-,2_\f^+,3^+,\Kh_{45}^+)
A_3^\tree((-\Kh_{45})^{-},4^+,\hat 5^+)\nonumber\\
&=& -{i\over2}{1\over s_{45}} 
 {\spash{\hat 1}.2\spbsh2.{\Kh_{45}}\over\spa2.3\spash3.{\Kh_{45}}}
 {\spbsh4.{\hat 5}^3\over \spbsh{(-\Kh_{45})}.4\spbsh{\hat 5}.{(-\Kh_{45})}}
\nonumber\\
&=& {i\over2}{1\over s_{45}} 
 { \spa1.2\sand1.{\Ksl_{45}}.2
  \over \spa2.3\sand3.{\Ksl_{45}}.5 }
 { \spb4.5^3 
  \over \sand1.{\Ksl_{45}}.4 }
\label{FirstTermRecurrence}\\
&=& {i\over2}{1 \over s_{45}} 
 {\spa1.2\spa1.3\spb3.2
  \over\spa2.3\spa3.4\spb4.5}
 { {\spb4.5}^2
  \over \spa1.5 }\nonumber\\
&=& {i\over2}
 {\spa1.2\spa1.3\spb2.3
  \over\spa2.3\spa3.4\spa4.5\spa1.5} \,,\nonumber
\end{eqnarray}
exactly the {\it first\/} term in $A_5^{L-s}$.  But the second term 
in \eqn{A5decomposition} is missing.

What happened to it?  Notice that the second term in~\eqn{A5decomposition}
is {\it not\/} singular in the $k_4\parallel k_5$ limit, so long as 
we are considering real momenta, because the numerator and denominator
vanish at the same rate in the limit.  If we consider {\it complex\/}
momenta, however, the behavior of $\spa4.5$ is decoupled from that
of $\spb4.5$, and in regions where $\spa4.5\rightarrow 0$, there is
a pole.

In other words: while multiparticle
and collinear factorization capture the full pole structure for 
{\it real\/} momenta, they do {\it not\/} do so for {\it complex\/}
momenta.  In particular, a term containing the unreal pole 
(\ref{UnrealPole})  in the $a\parallel b$ limit, 
will not contribute to a collinear singularity, but will contribute
to a pole for complex momenta (which decouple the behavior of $\spa{a}.b$
from that of $\spb{a}.b$).  That is, for our purposes
we need to ask what the quasi-universal
behavior in the collinear limit is of finite terms with non-trivial 
phase structure. 
For  real momenta the ratio (\ref{UnrealPole}) 
is always nonsingular, because 
$|\spa{a}.{b}| = |\spb{a}.{b}| = \sqrt{|2 k_{a}\cdot k_{b}|}$. (It
does contain a phase dependence, though, which selects it out uniquely 
in the collinear limit: as the vectors $\vec k_{a}$ and 
$\vec k_{b}$ are rotated around each other by an angle $\phi$ 
the ratio (\ref{UnrealPole}) will change by the phase factor $\exp(-2i\phi)$.)

%
\begin{figure}[t]
\centerline{\epsfxsize 1.8 truein \epsfbox{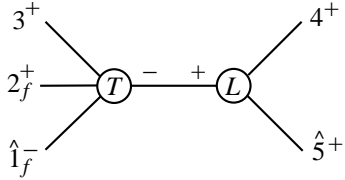}}
\caption{The extra unreal pole contribution in the recursion relation for 
$A_5^{L-s}(1_\f^+, 2_\f^+,3^+,4^+,5^+)$.}
\label{A5Lms15ExtraFigure}
\end{figure}

At tree level, a diagrammatic analysis that isolates the collinear
singularities for real momenta extends readily (for massless
amplitudes) to complex momenta.  However, the same is not true at loop level.
We will not offer such an analysis in this paper.  Rather, we will
exhibit a set of ans\"atze which use lower-point amplitudes and yield
the contributions required for the corresponding poles.  

We proceed as in ref.~\cite{OnShellRecurrenceI}.  Our experience there
suggests that the missing term is associated with a factorization of
the type shown in \fig{A5Lms15ExtraFigure}.  In order to define the 
3-point loop vertex in this diagram, we could try to use the factor 
$V_3^{\oneloop}$ vertex~(\ref{OneLoopThreeVertex}) which appeared in 
the gluonic case~\cite{OnShellRecurrenceI}.  However, \eqn{A5decomposition} 
contains a normalization factor of $1/2$ rather than $1/3$,
and is not associated directly with a double pole.
So we introduce the ``$L\!-\!s$ loop vertex'',
\begin{equation}
V_3^{L-s}(1^+,2^+,3^+) \equiv 
   - {i \over 2} \spb1.2 \spb2.3 \spb3.1 \,,
\end{equation}
which is proportional to $V_3^{\oneloop}$.

In the present case, there is no real-pole contribution associated 
with the diagram in \fig{A5Lms15ExtraFigure};
it would correspond to an ``$A^{s\,}$'' contribution which we have
subtracted out in $A^{L-s}$. (Accordingly, for complex
momenta, the double pole contributions present in
ref.~\cite{OnShellRecurrenceI} are absent.)  However, unreal pole
contributions can arise.  There are a number of constraints that allow
us to find the precise form of the required term.  The unreal pole
is power-suppressed by a factor of $s_{45}$ compared to the diagram in
\fig{A5Lms15ExtraFigure}. By dimensional analysis we then need
additional factors to obtain the correct overall dimensions.  The
additional factors must be invariant under phase rotations of spinors
associated with all external states, and the intermediate state
$\hat{K}_{45}$.  

We can use universal multiplicative ``soft factors,''
which describe the insertion of a soft gluon $s$ between two hard
partons $a$ and $b$ in a color-ordered amplitude, to construct such a
multiplicative function.  The soft factors depend only on the helicity
of the soft gluon and are given by~\cite{TreeReview},
\begin{eqnarray}
\Soft^\tree( a,  s^+, b) & =& {\spa{a}.{b} \over \spa{a}.{s} \spa{s}.{b}} \,,\\
\Soft^\tree( a,  s^-, b) & =& -{\spb{a}.{b} \over \spb{a}.{s} \spb{s}.{b}} \,.
\label{softdef}
\end{eqnarray}
They are invariant under phase rotations of spinors associated with 
$a$ and $b$, but not $s$.  However, the product 
\begin{equation}
s_{45} \Soft^\tree(a,  s^+, b) \Soft^\tree(c,  (-s)^-, d) 
\label{trysoft}
\end{equation}
is dimensionless, invariant under phase rotations of
$a$, $b$, $c$, $d$ and $s$, and suppressed as $K_{45}^2 \to 0$,
for suitable choices of $a$, $b$, $c$, $d$ and $s$.

Since the leg $s$ appears in two factors in \eqn{trysoft} carrying opposite
helicity, it is natural to identify it with the on-shell intermediate
momentum $\hat{K}_{45}$.  Choosing $c=\hat{5}$ and $d=4$ produces the
desired collinear behavior $\propto \spa4.5$ for~\eqn{trysoft}.  A
little experimentation shows that here the legs
$2$ and $\hat 1$ should be identified with $a$ and $b$.
Multiplying \eqn{trysoft} with these assignments by the ``naive''
diagram in \fig{A5Lms15ExtraFigure},
\begin{equation}
{i\over s_{45}^2} 
A_{4}^\tree(\hat 1_\f^-,2_\f^+,3^+,\Kh_{45}^-)
   \, V_3^{L-s}((-\Kh_{45})^+,4^+,\hat 5^+)\,,
\end{equation}
gives us exactly the missing contribution,
\begin{equation}
{i\over2} 
{\spa1.4\spb4.5\over\spa2.3\spa3.4\spa4.5} \,.
\end{equation}

More generally, in the $n$-point on-shell recursion relation for
$A_n^{L-s}(j_\f^+)$, when we choose the shifted legs 
in \eqn{SpinorShift} to be $(k,l)=(1,n)$, we need to add an 
unreal-pole term of the form,
\begin{eqnarray}
&&{i\over s_{(n-1)n}} 
A_{n-1}^\tree(\hat 1_\f^-,2^+,\ldots,j_\f^+,\ldots,(n-2)^+,\Kh_{(n-1) n}^-)
                    V_3^{L-s}((-\Kh_{(n-1)n})^+,(n-1)^+,\hat n^+) 
\nonumber\\ &&\hskip 15mm \null \times
\Soft^\tree(j, \Kh_{(n-1)n}^+,\hat 1) \, 
\Soft^\tree(\hat n, (-\Kh_{(n-1)n})^-, n-1) \,.
\label{DeltaExtraTerm}
\end{eqnarray}
The complete recursion relation for this choice of shifted legs is then,
\begin{eqnarray}
 && \null \hskip -10 mm A_n^{L-s}(j_\f^+) 
\nonumber\\
 &=& {i\over s_{(n-1)n}} 
A_{n-1}^{L-s}(\hat 1_\f^-,2^+,\ldots,j_\f^+,\ldots,(n-2)^+,\Kh_{(n-1) n}^+)
                A_3^{\tree}((-\Kh_{(n-1)n})^-,(n-1)^+,\hat n^+) 
\nonumber\\ 
&&\hskip 2mm \null + {i\over s_{(n-1)n}} 
A_{n-1}^\tree(\hat 1_\f^-,2^+,\ldots,j_\f^+,\ldots,(n-2)^+,\Kh_{(n-1) n}^-)
                V_3^{L-s}((-\Kh_{(n-1)n})^+,(n-1)^+,\hat n^+) 
\nonumber\\ &&\hskip 20mm  \null\times
\Soft^\tree(j, \Kh_{(n-1)n}^+,\hat 1) \,
\Soft^\tree(\hat n, (-\Kh_{(n-1)n})^-, n-1) \,,
\label{DeltaFullRecurrence}
\end{eqnarray}
where the hatted momenta in \eqn{DeltaFullRecurrence} are defined
by the shift
\begin{eqnarray}
&\tlambda_1 &\rightarrow \tlambda_1 - z\tlambda_n \,, \nonumber\\
&\lambda_n &\rightarrow \lambda_n + z\lambda_1 \,,
\label{SpinorShift1n}
\end{eqnarray}
with
\begin{equation}
 z = - { K_{n-1,n}^2 \over \spab1.{\Ksl_{n-1,n}}.n } 
  = - { \spa{(n-1)}.{n} \over \spa{(n-1)}.{1} } \,,
\end{equation}
in each term.  This relation assumes that, as in the case $n=5$, the
shifted amplitude $A(z)$ vanishes as $z\rightarrow \infty$, so there
is no surface term to be added to \eqn{NoSurface}.  The recursion
relation~(\ref{DeltaFullRecurrence}) can be solved, yielding the
following simple expression,
\begin{equation}
A_n^{L-s}(1_\f^-,2^+,\ldots,j_\f^+,\ldots,n^+) =
{i \over 2}
{\spa1.j \Sigma_{l=3}^{n-1} \sandmp1.{\Ksl_{2\cdots l}\ksl_l}.1
 \over \spa1.2\spa2.3\cdots\spa{n}.1} \,.
\label{allnLminuss}
\end{equation}
The reader may verify that this result has the correct collinear
limits in all channels, and that it obeys the reflection
symmetry~(\ref{Lminussflip}).

\subsection{Structure of Recursion for $s$ Contribution}

We turn next to the computation of the scalar contributions, given
by the $A_n^s$ function.  In this case, there is a non-trivial collinear
limit when $k_{n-1}\parallel k_n$ with real momenta, so that we expect
to have both double-pole and single-pole contributions with complex
momenta.  The latter may have an interpretation as `underlying' the
double-pole term, or else as simply being unreal poles.  The presence
of both types of term is similar to the gluon amplitudes with one
negative-helicity gluon considered in ref.~\cite{OnShellRecurrenceI}.  We can
again find an appropriate function using soft factors; the factors here are
however different from those in the gluon case.  As above, take the shifts
to be $(1,n)$.  In the five-point case, for example, the factors are,
\begin{equation}
\Soft^\tree(3,\Kh_{45}^+,2) \, \Soft^\tree(\hat 5,(-\Kh_{45})^-, 4) \,,
\end{equation}
when the positive-helicity quark is leg $2$, and absent if it is leg $3$.
This observation suggests a general form,
\begin{equation}
\Soft^\tree(n-2,\Kh_{(n-1)n}^+,j)\,\Soft^\tree(\hat n,(-\Kh_{(n-1)n})^-,n-1)\,,
\label{SExtraTerm}
\end{equation}
where $j$ labels the positive-helicity quark.

%
\begin{figure}[t]
\centerline{\epsfxsize 6. truein \epsfbox{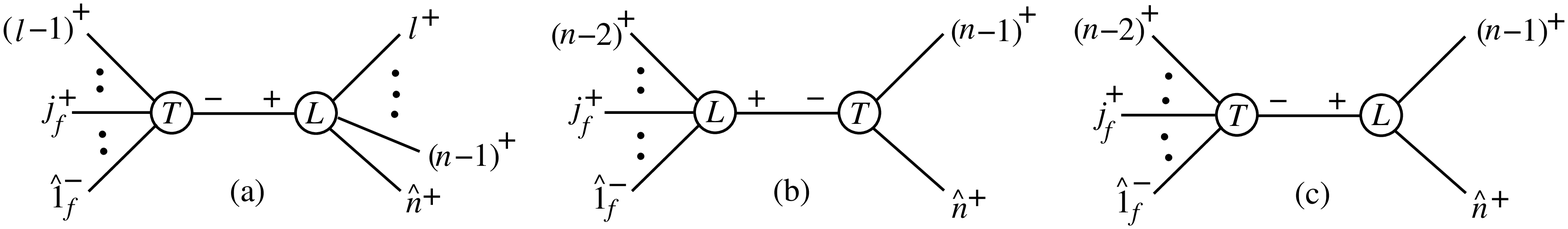}}
\caption{The diagrams corresponding to the terms in the recursion 
relation in \eqn{SFullRecurrence}. In diagram (a) $l$ runs over 
$\{j+1,j+2, \ldots, n-2\}$.  Diagram (c) contains a double pole
as well as an unreal pole underneath it.}
\label{An1sFigure}
\end{figure}

The full recursion relation for $A_n^s(j_\f^+)$, again assuming the
absence of a $z\rightarrow \infty$ surface term, is depicted in
\fig{An1sFigure}, and reads,
\begin{eqnarray}
&&\null \hskip -5 mm  A_n^s(j_\f^+) \nonumber \\
&& =\sum_{l=j+1}^{n-2} 
{i\over s_{l\ldots n}} 
A_{l}^\tree(1_\f^-,2^+,\ldots,j_\f^+,\ldots,(l-1)^+,\Kh_{l\ldots n}^-)
A_{n-l+2}^{\oneloop}((-\Kh_{l\ldots n})^+,l^+,\ldots,\hat n^+)\nonumber\\
&& \null +{i\over s_{(n-1)n}} 
A_{n-1}^{s}(\hat 1_\f^-,2^+,\ldots,j_\f^+,\ldots,(n-2)^+,\Kh_{(n-1) n}^+)
                A_3^{\tree}((-\Kh_{(n-1)n})^-,(n-1)^+,\hat n^+) 
\nonumber\\ 
&&\null +{i\over s_{(n-1)n}^2} 
A_{n-1}^\tree(\hat 1_\f^-,2^+,\ldots,j_\f^+,\ldots,(n-2)^+,\Kh_{(n-1) n}^-)
                V_3^{\oneloop}((-\Kh_{(n-1)n})^+,(n-1)^+,\hat n^+) 
\nonumber\\ &&\hskip 5mm \null\times
\Bigl(1 + s_{(n-1)n}\, \Soft^\tree(n-2, \Kh_{(n-1)n}^+,j) \,
\Soft^\tree(\hat n, (-\Kh_{(n-1)n})^-, n-1)
\Bigr) \,.
\label{SFullRecurrence}
\end{eqnarray}
The hatted legs undergo the shift in \eqn{SpinorShift1n}, where
in the $s_{l \ldots n}$ channel we set,
\begin{equation}
z = - {s_{l \ldots n} \over \sand1.{\Ksl_{l \ldots n}}.n} \,.
\end{equation}

The tree-side soft factor here, $\Soft^\tree(n-2, \Kh_{(n-1)n}^+,j)$,
is in a sense the ``complement'' of that for $A_n^{L-s}$ appearing
in~\eqn{DeltaExtraTerm}; that is,
\begin{equation}
\Soft^\tree(n-2, \Kh_{(n-1)n}^+,j)+
\Soft^\tree(j, \Kh_{(n-1)n}^+,\hat 1) = 
\Soft^\tree(n-2, \Kh_{(n-1)n}^+,\hat 1) \,,
\end{equation}
where the right-hand side is exactly the tree-side soft factor that 
appeared in our previous recursion relation~\cite{OnShellRecurrenceI} 
for the one-loop all-gluon amplitude with one negative helicity.

\subsection{Solution to $A_n^s$ Recursion Relation}

%
\begin{figure}[t]
\centerline{\epsfxsize 3.8 truein \epsfbox{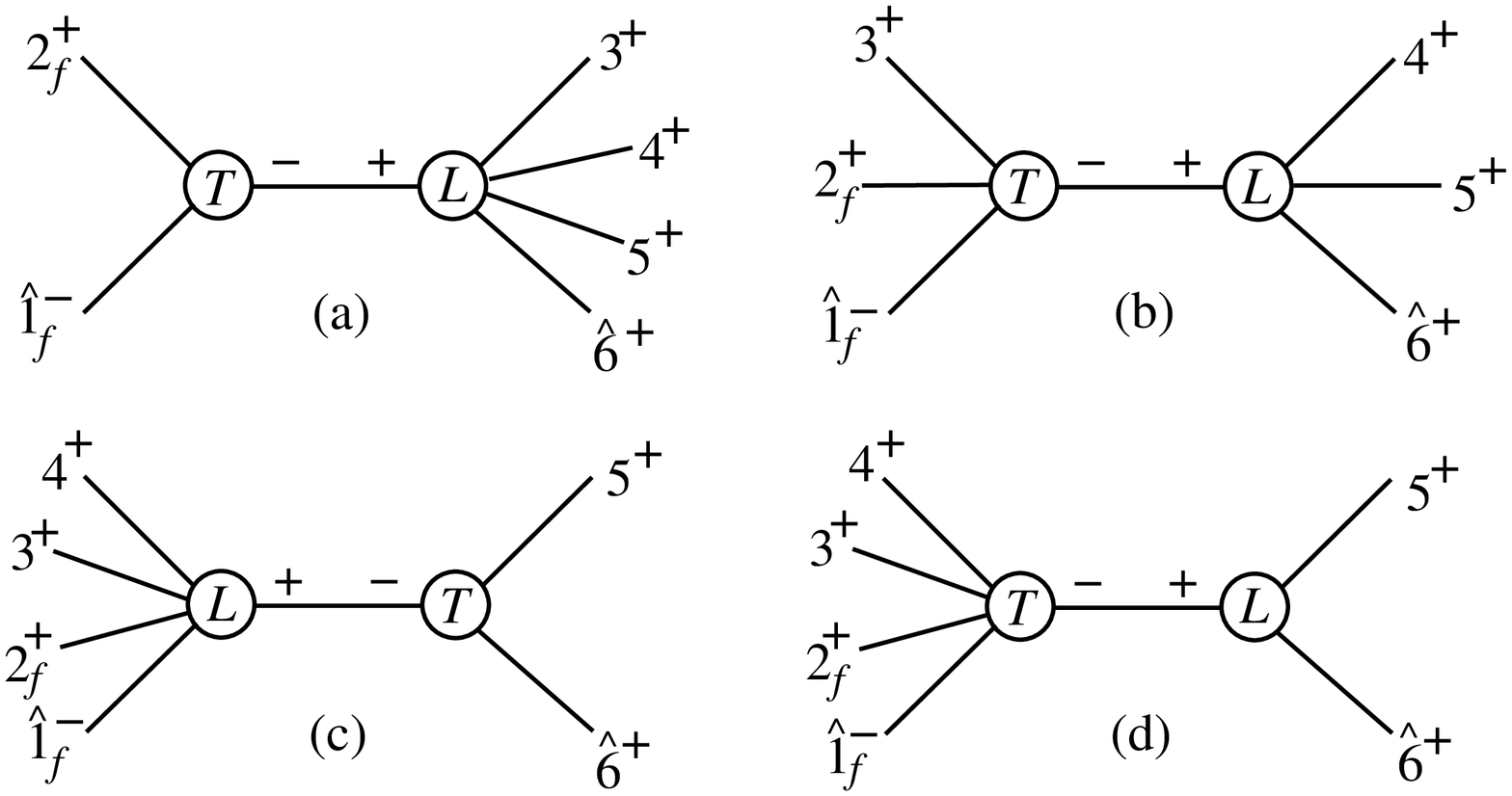}}
\caption{The recursive diagrams for $A_6^{s}(2_\f^+)$ using the shift
in \eqn{SpinorShift1n} with $n=6$.}
\label{A61sFigure}
\end{figure}

Unlike $A_n^{L-s}$, $A_n^s$ does have multi-particle poles.  Consider
the six-point case.  If we use the shift (\ref{SpinorShift1n}) the
recursion relation~(\ref{SFullRecurrence}) generates the diagrams shown in
\fig{A61sFigure} for the case where leg 2 is the positive
helicity fermion leg. The other cases, where the positive-helicity fermion
is leg~3 or~4, are similar.  Evaluating these contributions, we find,
\begin{eqnarray}
A_6^{s}(2_\f^+) &=& 
{i\over3} \Biggl(
  { \spa1.4 \sandmp1.{(2+3)(3+4)}.1
  \over \spa1.2 {\spa3.4}^2 \spa4.5 \spa5.6 \spa6.1 }
+ { \spa2.4 \spa1.5 \sandmp1.{(4+5)(5+6)}.1
   \over \spa1.2 \spa2.3 \spa3.4 {\spa4.5}^2 \spa5.6 \spa6.1 }
\nonumber\hskip 1.5 cm \\ && \hskip5mm
- { \spa2.5 \sandmp1.{5\,6}.1
   \over \spa1.2 \spa2.3 \spa3.4 \spa4.5 {\spa5.6}^2 }
+ { {\spab1.{(3+4)}.2}^2 
   \over {\spa3.4}^2 \spa5.6 \spa6.1 \spab5.{(3+4)}.2 }
\nonumber\\ && \hskip5mm
- { \spb2.6^2 \sandpm2.{(3+4)(4+5)(3+4)(4+5)}.6
  \over \spb1.2 \spa3.4 \spa4.5 
  \spab5.{(3+4)}.2 \spab3.{(4+5)}.6 s_{345} }
\nonumber\\ && \hskip5mm 
+ { \spab2.{(4+5)}.6 {\spab1.{(4+5)}.6}^2 
   \over \spa1.2 \spa2.3 {\spa4.5}^2 
   \spab3.{(4+5)}.6 s_{456} }
  \Biggr) \,,
\label{A62s}\\
A_6^{s}(3_\f^+) &=&
{i\over3} \Biggl(
  {\spa1.5\spa1.6\spa3.5\spb5.6\over \spa1.2\spa2.3\spa3.4\spa4.5\spa5.6^2}
 + {\spa1.5\sandmp1.{(4+5)(5+6)}.1\over
   \spa1.2\spa2.3\spa4.5^2\spa5.6\spa6.1}
\nonumber\\ && \hskip5mm 
 + {\sand1.{(4+5)}.6^2\over\spa1.2\spa2.3\spa4.5^2 s_{123}}
  \Biggr),\label{A63s}\\
A_6^{s}(4_\f^+) &=& 
- {i\over3}
  {\sandmp1.{5\,6}.1 \over\spa1.2\spa2.3\spa3.4\spa5.6^2} \,.
\label{A64s}
\end{eqnarray}
These expressions suggest the all-$n$ forms,
\begin{eqnarray}
A_n^{s}((n-2)_\f^+) &=&
 -{i\over 3}{\sandmp1.{(n-1)\,n}.1 \over
  \spa1.2\cdots\spa{(n-3)}.{(n-2)}\spa{(n-1)}.n^2}
,~~~~~\label{nm2falln}\\
A_n^{s}((n-3)_\f^+) &=&
{i\over3}\Biggl(
- {\sandmp1.{(n-1)\,n}.1 \spa{(n-3)}.{(n-1)}
    \over\spa1.2\spa2.3\cdots \spa{(n-2)}.{(n-1)}\spa{(n-1)}.n^2}\nonumber\\
&&\hphantom{{i\over3}\Biggl()\!}
 + {\spa1.{(n-1)} \sandmp1.{\Ksl_{n-2,n-1} \Ksl_{n-1,n}}.1
    \over \spa1.2\spa2.3\cdots  \spa{(n-4)}.{(n-3)}
          \spa{(n-2)}.{(n-1)}^2\spa{(n-1)}.n\spa{n}.1}\nonumber\\
&&\hphantom{{i\over3}\Biggl()\!}
+ {\sand1.{\Ksl_{n-2,n-1}}.n^2
    \over\spa1.2\spa2.3\cdots \spa{(n-4)}.{(n-3)}
          \spa{(n-2)}.{(n-1)}^2 s_{(n-2)\cdots n} }\Biggr) \,.
\label{nm3falln}
\end{eqnarray}

By studying the structure of eqs.~(\ref{A62s})--(\ref{nm3falln}), and
also the output of the recursion relation for $n=7$ and $n=8$, we have
arrived at the following compact formula, valid for all $j$ and $n$,
\begin{equation}
A_n^{s}(j_\f^+) =
{i\over3} 
  {S_1 + S_2 \over \spa1.2\spa2.3\cdots \spa{n}.{1} } \,,
\label{alljn}
\end{equation}
where
\begin{eqnarray}
 S_1 &=& \sum_{l=j+1}^{n-1} 
  { \spa{j}.{l} \spa{1}.{(l+1)} 
    \sandmp1.{\Ksl_{l,l+1} \Ksl_{(l+1)\cdots n}}.1
   \over \spa{l}.{(l+1)} } \,,
\label{alljnS1} \\
 S_2 &=& \sum_{l=j+1}^{n-2} \sum_{p=l+1}^{n-1}
 { \spa{(l-1)}.{l}
   \over \sandmp1.{\Ksl_{(p+1)\cdots n} \Ksl_{l\cdots p}}.{(l-1)}
         \sandmp1.{\Ksl_{(p+1)\cdots n} \Ksl_{l\cdots p}}.{l} }    
\nonumber \\
&& \hskip15mm\times 
 { \spa{p}.{(p+1)}
   \over \sandmp1.{\Ksl_{2\cdots (l-1)} \Ksl_{l\cdots p}}.{p}
         \sandmp1.{\Ksl_{2\cdots (l-1)} \Ksl_{l\cdots p}}.{(p+1)} }
\nonumber \\
&& \hskip15mm\times
   {\sandmp1.{\Ksl_{l\cdots p} \Ksl_{(p+1)\cdots n}}.1}^2
       \sandmp{j}.{\Ksl_{l\cdots p} \Ksl_{(p+1)\cdots n}}.1 
\nonumber \\
&& \hskip15mm\times
   { \sandmp1.{\Ksl_{2\cdots (l-1)} [ {\cal F}(l,p) ]^2  \Ksl_{(p+1)\cdots n}}.1
    \over s_{l\cdots p} }
 \,,
\label{alljnS2}
\end{eqnarray}
and 
\begin{equation}
{\cal F}(l,p) = \sum_{i=l}^{p-1} \sum_{m=i+1}^{p} \ksl_i \ksl_m \,.
\label{Flpdef}
\end{equation}
%


\subsection{Verification of Solution}
\label{SolVerificationSubSection}

We now verify analytically that these amplitudes
satisfy the recursion relation~(\ref{SFullRecurrence}).
First consider the term shown in \fig{An1sFigure}(b),
containing the three-point tree amplitude
$A_3^{\tree}((-\Kh_{(n-1)n})^-,(n-1)^+,\hat n^+)$.
Let $\hat{S}_1$ and $\hat{S}_2$ stand for
the shifted versions of $S_1$ and $S_2$ for the appropriate 
$(n-1)$-point one-loop quark amplitude.
The term in \fig{An1sFigure}(b) can be simplified to 
\begin{eqnarray}
&&
- {1\over K_{n-1,n}^2} 
{ {\spb{(n-1)}.{n}}^3 
  \over \spbsh{n}.{\hat{K}_{n-1,n}} \spbsh{\hat{K}_{n-1,n}}.{(n-1)} }
\nonumber\\
&&\hskip12mm \times
 {i\over3} { \hat{S}_1 + \hat{S}_2 
\over \spash{\hat{K}_{n-1,n}}.{1} \spa1.2 \spa2.3 \cdots 
         \spa{(n-3)}.{(n-2)} \spash{(n-2)}.{\hat{K}_{n-1,n}} }
\nonumber \\
&=& {i\over3} { \hat{S}_1 + \hat{S}_2 \over \spa1.2 \spa2.3 \cdots \spa{n}.1 }
{ \spa{(n-2)}.{(n-1)} \spa{(n-1)}.{n} \spa{n}.1 
  {\spb{(n-1)}.{n}}^3
  \over \spa{(n-1)}.{n} \spb{(n-1)}.{n} 
  \spab1.{\Ksl_{n-1,n}}.{{(n-1)}} \spab{(n-2)}.{\Ksl_{n-1,n}}.{n} }
\nonumber\\
&=&  {i\over3} { \hat{S}_1 + \hat{S}_2  
       \over \spa1.2 \spa2.3 \cdots \spa{n}.1 } \,.
\label{PrefactorBehavior}
\end{eqnarray}
We see that the correct spinor denominator factor is reproduced.

The next question is how $\hat{S}_1$ and $\hat{S}_2$ are
affected by the shift~(\ref{SpinorShift1n}).
Because $\hat{S}_1$ is simpler, we discuss it first.
Note that the $(n-1)$-point expression $\hat{S}_1$ is a 
single sum over $l$ containing $(n-2)-j$ terms, which is 
one fewer than the number of terms in the $n$-point 
expression $S_1$ we are trying to produce.
All terms but the last in $\hat{S}_1$ have very simple behavior
under the shift~(\ref{SpinorShift1n}) of $\tlambda_1$ and 
$\lambda_n$.  They depend on $\lambda_1$ through $\langle 1^-|$ 
and $|1^+\rangle$, but do not depend on $\tlambda_1$.  
Their dependence on $\lambda_n$ is solely via the factor
\begin{equation}
\ldots (\Ksl_{(l+1)\cdots (n-2)} + \hat{\Ksl}_{n-1,n})  | 1^+\rangle
= \ldots \Ksl_{(l+1)\cdots n} | 1^+ \rangle \,,
\label{spec1}
\end{equation}
because the shift in $\Ksl_{n-1,n}$ is proportional to $\lambda_1$.
Thus each such term directly yields the corresponding 
term in the $n$-point sum $S_1$.  

The last term in $\hat{S}_1$,
with $l=n-2$, is the exception, because it has additional dependence
on $\Ksl_{n-1,n}$.  It can be written as
\begin{eqnarray}
&&
{ \spa{j}.{(n-2)} \spash1.{\hat{K}_{n-1,n}} \spbsh{\hat{K}_{n-1,n}}.{n}
   \sandmp1.{(\ksl_{n-2} + \Ksl_{n-1,n}) \Ksl_{n-1,n}}.1
  \over \spash{(n-2)}.{\hat{K}_{n-1,n}}
        \spbsh{\hat K_{n-1,n}}.{n} }
\nonumber\\
&=& { \spa{j}.{(n-2)} \spa1.{{(n-1)}} 
   \sandmp1.{\ksl_{n-2} \Ksl_{n-1,n}}.1
  \over \spa{(n-2)}.{(n-1)} } \,,
\label{LastS1hatterm}
\end{eqnarray}
which does not quite match the next-to-last term ($l=n-2$) 
in $S_1$,
\begin{equation}
{ \spa{j}.{(n-2)} \spa1.{{(n-1)}} 
   \sandmp1.{\Ksl_{n-2,n-1} \Ksl_{n-1,n}}.1
  \over \spa{(n-2)}.{(n-1)} } \,.
\label{NexttoLastS1term}
\end{equation}

Additional contributions to the next-to-last term, and the whole
of the last term ($l=n-1$) in $S_1$, come from the
term containing $V_3^{\oneloop}((-\Kh_{(n-1)n})^+,(n-1)^+,\hat n^+)$
in the recursion relation~(\ref{SFullRecurrence}), 
depicted in \fig{An1sFigure}(c).
This term can be simplified to,
\begin{eqnarray}
&&
{i\over3} {1\over s_{(n-1)n}^2} 
{ {\spash{1}.{\hat K_{n-1,n}}}^2 \spash{j}.{\hat{K}_{n-1,n}}
  \spbsh{\hat{K}_{n-1,n}}.{(n-1)} \spb{(n-1)}.{n} \spbsh{n}.{\hat K_{n-1,n}} 
  \over \spa1.2 \spa2.3 \cdots \spa{(n-3)}.{(n-2)} 
  \spash{(n-2)}.{\hat{K}_{n-1,n}} }
\nonumber \\
&& \hskip7mm
\times \Biggl( 1 
 + s_{(n-1)n} 
   { \spa{(n-2)}.{j} \spb{n}.{(n-1)}
     \over \spash{(n-2)}.{\hat{K}_{n-1,n}} \spash{\hat{K}_{n-1,n}}.{j}
      \spbsh{n}.{\hat K_{n-1,n}} \spbsh{\hat{K}_{n-1,n}}.{(n-1)}  } \Biggr)
\nonumber\\
&=&
- {i\over3} {1 \over \spa1.2\spa2.3\cdots \spa{(n-3)}.{(n-2)} }
\nonumber\\
&& \hskip7mm
\times
   { \spab1.{\Ksl_{n-1,n}}.{{(n-1)}} \spab1.{\Ksl_{n-1,n}}.{n}  
     \spab{j}.{\Ksl_{n-1,n}}.{n} \spb{(n-1)}.{n} 
    \over {\spa{(n-1)}.{n}}^2 {\spb{(n-1)}.{n}}^2
    \spab{(n-2)}.{\Ksl_{n-1,n}}.{n} }
\nonumber\\
&& \hskip7mm
\times \Biggl( 1 
    - { \spa{j}.{(n-2)} \spa{(n-1)}.{n} {\spb{(n-1)}.{n}}^2
       \spab{1}.{\Ksl_{n-1,n}}.{n} 
      \over \spab{(n-2)}.{\Ksl_{n-1,n}}.{n} 
            \spab{j}.{\Ksl_{n-1,n}}.{n} 
            \spab{1}.{\Ksl_{n-1,n}}.{{(n-1)}} } \Biggr)
\nonumber\\
&=&
{i\over3} {1\over \spa1.2\spa2.3\cdots \spa{n}.1 }
   { \spa{j}.{(n-1)} \spa1.{n} \sandmp1.{\Ksl_{n-1,n} \ksl_n}.1
   \over \spa{(n-1)}.{n} }
\nonumber\\
&& \hskip7mm
\times \Biggl( 1 
    + { \spa{j}.{(n-2)} \spa{(n-1)}.{n} \spa{1}.{(n-1)}
      \over \spa{j}.{(n-1)} \spa{(n-2)}.{(n-1)} \spa{1}.{n} } \Biggr)
\nonumber\\
&=&
{i\over3} {1\over \spa1.2\spa2.3\cdots \spa{n}.1 }
 \Biggl(
   { \spa{j}.{(n-1)} \spa1.{n} \sandmp1.{\Ksl_{n-1,n} \ksl_n}.1
   \over \spa{(n-1)}.{n} }
\nonumber\\
&& \hskip36mm
 + { \spa{j}.{(n-2)} \spa1.{(n-1)} \sandmp1.{\ksl_{n-1} \Ksl_{n-1,n}}.1
   \over \spa{(n-2)}.{(n-1)} }
  \Biggr) \,.
\label{SoftTermsS1}
\end{eqnarray}
The term containing $\spa{j}.{(n-2)}$ in \eqn{SoftTermsS1} combines
with \eqn{LastS1hatterm} to generate the next-to-last 
term~(\ref{NexttoLastS1term}) in $S_1$, while the term 
containing $\spa{j}.{(n-1)}$ is just the last term in $S_1$.
So we have demonstrated that all of the terms in $S_1$
are produced correctly by the recursion relation.

Now we turn to $S_2$.  We divide the $S_2$ terms into those
with $p<n-1$ and those with $p=n-1$.  The bulk of the terms, 
with $p<n-1$, come from the $\hat{S}_2$ contribution we left
unexamined in \eqn{PrefactorBehavior}.  To show this,
we observe that (as was the case for the bulk of the $\hat{S}_1$
terms) $\tlambda_1$ never appears in $\hat{S}_2$. 
Also, $k_n$ only appears in \eqn{alljnS2} for $S_2$ via 
$\langle 1^- | \Ksl_{(p+1)\cdots n} \ldots$
or $\ldots\Ksl_{(p+1)\cdots n} | 1^+\rangle$.
Thus we may again apply \eqn{spec1} (with $l$ replaced by $p$)
to $\hat{S}_2$, in order to see that every term with $p<n-1$ 
in $S_2$ is generated directly from the corresponding term 
in $\hat{S}_2$.

The $S_2$ terms with $p=n-1$ come from the terms containing
the one-loop pure-glue all-plus amplitude $A_{n-l+2}^{\oneloop}$
in \eqn{SFullRecurrence}, depicted in \fig{An1sFigure}(a).  
The $l^{\rm th}$ term in $S_2$ comes from the $l^{\rm th}$ term in 
the recursion relation.  The main technical detail is to 
rewrite the numerator $H_n$ of the all-plus amplitude,
as given in \eqn{Hndef}, so that it looks like the operator
$[{\cal F}(l,n-1)]^2$ appearing in the terms with $p=n-1$ in $S_2$.
For this purpose, we first use the identity~(\ref{FFtoshow2}), 
with $p=n$, to rewrite the numerator factor $H_{n-l+2}(l,n)$  
(for the all-plus amplitude with gluon momenta 
$k_l, k_{l+1}, \ldots, k_n, P$, where $P = - K_{l\cdots n}$ is on-shell) 
as,
\begin{equation}
H_{n-l+2}(l,n) = - {1\over2} \Tr_{-} [ {\cal F}(l,n) ]^2 \,.
\label{HtoTrF}
\end{equation}
Then we use the identity
\begin{equation}
 {1\over2} \Tr_{-} [ {\cal F}(l,n) ]^2 
 = { 
 \langle n^+ | \, [\hat{{\cal F}}(l,n)]^2 \, 
 | \hat{K}_{l\cdots n}^- \rangle
    \over \spbsh{n}.{\hat{K}_{l\cdots n}} } \,,
\label{Fspinor2}
\end{equation}
where $\hat{{\cal F}}$ is defined in \eqn{hatFlpdef},
which follows similarly from \eqns{FFspinoreq}{XFY} for $p=n$.
Finally, \eqn{Fextend} shows that we can shift $n \to n-1$
in the argument of $\hat{{\cal F}}$ in \eqn{Fspinor2}.
Combining this equation with \eqn{HtoTrF}, reversing
the order of the spinor strings, using momentum conservation,
and multiplying numerator and denominator by common factors, 
we arrive at the identity,
\begin{equation}
H_{n-l+2}(l,n) = 
- { \sandmp1.{\Ksl_{2\cdots(l-1)} [ {\cal F}(l,n-1) ]^2 \ksl_n}.1
  \over \sandmp1.{\Ksl_{2\cdots(l-1)} \ksl_n}.1 } \,.
\label{finalHtoFeq}
\end{equation}

A few other identities are also useful,
\begin{eqnarray}
\spa{n}.{1} \, s_{l\cdots n} &=& 
\sandmp1.{\Ksl_{2\cdots(l-1)} \Ksl_{l\cdots(n-1)}}.{n} \,,
\label{moreid1}\\
\spab{\hat{n}}.{\Ksl_{l\cdots n}}.{n} &=&
\spab{n}.{\Ksl_{l\cdots n}}.{n} 
- { K^2_{l\cdots n} \over \spab1.{\Ksl_{l\cdots n}}.{n} }
  \spab1.{\Ksl_{l\cdots n}}.{n} 
\nonumber\\
&=& - s_{l\cdots(n-1)} \,,
\label{moreid2}\\
\spa{(n-1)}.{\hat{n}} &=&
\spa{(n-1)}.{n}
- { K^2_{l\cdots n} \over \spab1.{\Ksl_{l\cdots n}}.{n} }
  \spa{(n-1)}.{1}
\nonumber\\
&=& - { \sandmp1.{\Ksl_{2\cdots(l-1)} \Ksl_{l\cdots(n-1)}}.{{(n-1)}} 
        \over \spab1.{\Ksl_{l\cdots n}}.{n} } \,.
\label{moreid3}
\end{eqnarray}
Using these identities, the $l^{\rm th}$ term in the recursion relation can
be written as,
\begin{eqnarray}
&&
-{i\over3} {1 \over s_{l\cdots n}} 
 { {\spash1.{\hat K_{l\cdots n}}}^2 \spash{j}.{\hat{K}_{l\cdots n}}
  \over \spa1.2 \spa2.3 \cdots \spa{(l-2)}.{(l-1)} 
   \spash{(l-1)}.{\hat{K}_{l\cdots n}} }
\nonumber\\
&& \hskip1cm
\null \times
 { H_{n-l+2}(l,n) 
  \over \spash{\hat{K}_{l\cdots n}}.{l} \spa{l}.{(l+1)} \cdots 
   \spa{(n-1)}.{\hat{n}} \spash{\hat{n}}.{\hat{K}_{l\cdots n}} }
\nonumber\\
&=& - {i\over3} {1 \over \spa1.2 \spa2.3 \cdots \spa{n}.1}
 { \spa{(l-1)}.{l} \spa{(n-1)}.{n} \spa{n}.1
  {\spab1.{\Ksl_{l\cdots n}}.{n}}^2 \spab{j}.{\Ksl_{l\cdots n}}.{n}
  \over \spba{n}.{\Ksl_{l\cdots n}}.{{(l-1)}}
        \spba{n}.{\Ksl_{l\cdots n}}.{{l}} }
\nonumber\\
&& \hskip1cm
\null \times 
 { \spa{n}.1 
   \over \sandmp1.{\Ksl_{2\cdots(l-1)} \Ksl_{l\cdots(n-1)}}.{n} }
{ \spab1.{\Ksl_{l\cdots n}}.{n}
 \over \sandmp1.{\Ksl_{2\cdots(l-1)} \Ksl_{l\cdots(n-1)}}.{{(n-1)}} }
\nonumber\\
&& \hskip1cm
\null \times 
{1\over s_{l\cdots(n-1)}}
{ \sandmp1.{\Ksl_{2\cdots(l-1)} [ {\cal F}(l,n-1) ]^2 \ksl_n}.1
  \over \sandmp1.{\Ksl_{2\cdots(l-1)} \ksl_n}.1 } 
\nonumber\\
&=& {i\over3} {1 \over \spa1.2 \spa2.3 \cdots \spa{n}.1}
 { \spa{(l-1)}.{l}
   \over \sandmp1.{\ksl_n \Ksl_{l\cdots (n-1)}}.{(l-1)}
         \sandmp1.{\ksl_n \Ksl_{l\cdots (n-1)}}.{l} }    
\nonumber \\
&& \hskip1cm \null \times 
 { \spa{(n-1)}.{n}
   \over \sandmp1.{\Ksl_{2\cdots (l-1)} \Ksl_{l\cdots (n-1)}}.{{(n-1)}}
         \sandmp1.{\Ksl_{2\cdots (l-1)} \Ksl_{l\cdots (n-1)}}.{n} }
\nonumber \\
&& \hskip1cm \null \times
   {\sandmp1.{\Ksl_{l\cdots(n-1)} \ksl_n}.1}^2
    \sandmp{j}.{\Ksl_{l\cdots(n-1)} \ksl_n}.1 
\nonumber \\
&& \hskip1cm \null \times
   { \sandmp1.{\Ksl_{2\cdots (l-1)} [ {\cal F}(l,n-1) ]^2  \ksl_n}.1
    \over s_{l\cdots (n-1)} } \,.
\label{Finallthterm}
\end{eqnarray}
The final form is just the $l^{\rm th}$ term with $p=n-1$ in
$S_2$ in the solution~(\ref{alljn}).  This completes the demonstration
that \eqn{alljn} obeys the recursion 
relation~(\ref{SFullRecurrence}).

\subsection{Factorization Properties of Solution}
\label{FactorSolutionSubSection}

In \app{MultiFactAppendix} we show that formula~(\ref{alljn}) has all
the correct multi-particle poles, factorizing properly onto products
of the quark-containing MHV tree amplitudes~(\ref{ffmhvtree}) and the
one-loop all-plus-helicity pure-gluon
amplitudes~(\ref{OneLoopAllPlusAmplitude}).

The collinear singularities of \eqn{alljn} are also quite manifest,
although we shall not verify them all in detail here.  Most of the
factors that diverge in the collinear limits are contained in the
$1/(\spa1.2 \cdots \spa{n}.1)$ prefactor.  The generic collinear limit
factorizes onto another quark amplitude in the same 
sequence~(\ref{alljn}), but with one fewer external gluon.  
If the two particles
becoming collinear are color-adjacent gluons labeled $l$ and $l+1$,
with $l>j$, then the amplitude can also factorize
onto the product of the helicity-flip loop splitting amplitude,
$\Split^{\oneloop}_{+}(l^{+},(l+1)^{+};z) \propto \spb{l}.{(l+1)} /
{\spa{l}.{(l+1)}}^2$, and a quark-containing MHV tree
amplitude~(\ref{ffmhvtree}).  The corresponding terms in
$A_n^s(j_\f^+)$ come partly from the $l^{\rm th}$ term in $S_1$, and
partly from the term in $S_2$ with $p=l+1$.  The latter term has an
apparent singularity from a manifest factor of $1/s_{l,l+1}$.
However, this factor is cancelled by the numerator factor containing
$[ {\cal F}(l,l+1) ]^2 = \ksl_l \ksl_{l+1} \ksl_l \ksl_{l+1} =
s_{l,l+1} \times {\cal F}(l,l+1)$.  On the other hand, the denominator
factor $\sandmp1.{\Ksl_{2\cdots (l-1)} \Ksl_{l\cdots p}}.{p}$ produces
a spinor product $\spa{l}.{(l+1)}$, like the one manifest in the
$l^{\rm th}$ term in $S_1$.

When the positive-helicity fermion is in the position $j=2$, 
the limit $1 \parallel 2$ (where the anti-quark and quark momenta
become collinear) factorizes the finite quark amplitudes~(\ref{alljn}) 
onto the finite pure-glue amplitudes, according to \eqn{QuarkAntiQuarkLimit}. 
A term containing $1/\spb1.2$, corresponding to the factor
$\Split^\tree_{-}(1_\f^-,2_\f^+;z)$ in \eqn{QuarkAntiQuarkLimit},
will factorize onto the all-plus 
amplitudes~(\ref{OneLoopAllPlusAmplitude}).
The relevant term in $A_n^s(2_\f^+)$ is the term in $S_2$ with
$l=3$ and $p=n-1$.  The denominator factor of $\spb1.2$
comes from the string 
$\sandmp1.{\Ksl_{2\cdots (l-1)} \Ksl_{l\cdots p}}.{(p+1)}$
appearing in this term.
This limit can be checked by analysis very analogous 
to the discussion of the multi-particle factorization 
limit $K_{l\cdots n}^2 \to 0$ at the end of \app{MultiFactAppendix}.

The terms containing a factor of $1/\spa1.2$, corresponding to the 
factor $\Split^\tree_{+}(1_\f^-,2_\f^+;z)$ in \eqn{QuarkAntiQuarkLimit},
will factorize onto
all-gluon amplitudes with one negative-helicity gluon, 
$A_n^\oneloop(1^-,2^+,\ldots,n^+)$.  We can use
this factorization to extract a compact form for these pure-glue amplitudes.
  In this case, virtually every term in 
$S_1$ and $S_2$ contributes, except for those terms in $S_2$ 
with $l=3$.  (In each of the $l=3$ terms, the factor of $1/\spa1.2$ 
from the prefactor is cancelled by factors in the $S_2$ term itself.)
The compact form for the $n$-gluon amplitude that we derive from this 
limit is very similar to the quark amplitude itself (with $j \to 1$):
\begin{equation}
A_n^{\oneloop}(1^-,2^+,3^+,\ldots,n^+) =
{i\over3} 
  {T_1 + T_2 \over \spa1.2\spa2.3\cdots \spa{n}.{1} } \,,
\label{oneminusalln}
\end{equation}
where
\begin{eqnarray}
 T_1 &=& \sum_{l=2}^{n-1} 
  { \spa{1}.{l} \spa{1}.{(l+1)} 
    \sandmp1.{\Ksl_{l,l+1} \Ksl_{(l+1)\cdots n}}.1
   \over \spa{l}.{(l+1)} } \,,
\label{oneminusallnT1} \\
 T_2 &=& \sum_{l=3}^{n-2} \sum_{p=l+1}^{n-1}
 { \spa{(l-1)}.{l}
   \over \sandmp1.{\Ksl_{(p+1)\cdots n} \Ksl_{l\cdots p}}.{(l-1)}
         \sandmp1.{\Ksl_{(p+1)\cdots n} \Ksl_{l\cdots p}}.{l} }    
\nonumber \\
&& \hskip15mm\times 
 { \spa{p}.{(p+1)}
   \over \sandmp1.{\Ksl_{2\cdots (l-1)} \Ksl_{l\cdots p}}.{p}
         \sandmp1.{\Ksl_{2\cdots (l-1)} \Ksl_{l\cdots p}}.{(p+1)} }
\nonumber \\
&& \hskip15mm\times
   {\sandmp1.{\Ksl_{l\cdots p} \Ksl_{(p+1)\cdots n}}.1}^3
\nonumber \\
&& \hskip15mm\times
   { \sandmp1.{\Ksl_{2\cdots (l-1)} [ {\cal F}(l,p) ]^2  \Ksl_{(p+1)\cdots n}}.1
    \over s_{l\cdots p} }
 \,.
\label{oneminusallnT2}
\end{eqnarray}
This result agrees with eqs.~(\ref{mppp}), (\ref{mppppsimple})
and (\ref{mpppppsimple}), although it is written in a slightly different
form.  Furthermore, we have checked numerically that it agrees with 
the all-$n$ result of Mahlon~\cite{Mahlon} up through $n=18$.

\section{Further Cross Checks}
\label{CrossChecksSection}

There are a number of non-trivial checks we have performed on our
results.  As we already discussed, they have the correct factorization
properties in all collinear and multiparticle factorization channels
(with real momenta).  Another powerful check, which we describe now,
is that the amplitudes can be used to obtain formul\ae{} for
certain known results for QED and mixed QED/QCD amplitudes, 
and they agree with those earlier results.

We also comment on the consistency of using various shift
variables. In particular, there are choices of shift without the
subtlety of unreal poles, which lead to alternative recursion
relations.  These relations are also satisfied by the all-$n$
expressions~(\ref{allnLminuss}) and~(\ref{alljn}).

\subsection{Checks Based on QED Amplitudes}
\label{QEDCrossChecksSection}

Mahlon~\cite{MahlonQED,Mahlon} 
has computed the one-loop amplitudes for two separate processes,
which can be related to the ones presented here by converting
gluons into photons using appropriate permutation sums.  
These results therefore provide a stringent cross
check which is independent of the subtleties associated with unreal
poles.  A more general discussion of how to convert primitive QCD 
amplitudes with an external $q\bar{q}$ pair into QED amplitudes 
may be found in appendix D of ref.~\cite{TwoQuarkThreeGluon}.

In ref.~\cite{MahlonQED}, the pure QED amplitudes for
$e^+e^- \to \gamma^+ \gamma^+ \ldots \gamma^+$ were
computed.  In ref.~\cite{Mahlon}, the mixed QED/QCD 
amplitudes for $e^+e^- \to g^+ g^+ \ldots g^+$, and
for $\gamma^\pm g^+ g^+ \ldots g^+$, via a massless quark loop,
were presented.   Both sets of computations were performed using
a recursively-constructed tree-level current for two off-shell
massless fermions and an arbitrary number of gluons (or
photons)~\cite{MahlonQEDCurrent,Mahlon}.
To obtain the one-loop amplitudes,
this current is ``sewn up'' into a loop by joining the two off-shell
ends with the vertex for emission of a real photon, or of a virtual 
photon coupled to an electron-positron pair.

Consider first the case of $e^+e^- \to \gamma^+ \gamma^+ \ldots \gamma^+$.
Here there are two types of contributions, termed ${\cal A}_1$
and ${\cal A}_2$ in ref.~\cite{MahlonQED}.  The ${\cal A}_1$
contributions include a closed fermion loop, while the ${\cal A}_2$
contributions do not.  Both contributions allow for 
photons to be emitted off the external fermion line,
as well as from the closed fermion loop (in the case of ${\cal A}_1$).
The ${\cal A}_1$ piece is related to our $A^s$ amplitude, while
the ${\cal A}_2$ piece is related to $A^{L-s}$.

The ${\cal A}_2$ piece is a bit simpler and can be checked
analytically, so we begin with it.  Consider the primitive QCD
amplitude $A_n^L(n_f^+) \equiv A_n^L(1_f^-,2^+,\ldots,(n-1)^+,n_f^+)$.
According to \eqn{FLoopVanish}, there is no fermion or scalar (or
gluon loop) contribution in this case (simply because there are 
no external gluons on the same side of the external fermion line 
as the putative closed loop, and tadpole diagrams vanish here).  
Hence $A_n^s(n_f^+) = 0$, and 
$A_n^L(n_f^+) = A_n^{L-s}(n_f^+)$ 
is given by \eqn{allnLminuss} for $j=n$.
To convert this primitive amplitude into the QED amplitude 
${\cal A}_2$ with no closed fermion
loop, we merely need to sum \eqn{allnLminuss} for $j=n$ over all
$(n-2)!$ permutations of the $n-2$ gluons.  
The sum over permutations cancels the diagrams with gluon self-interactions, 
but retains abelian emission off the ``left'' side of the fermion line.
Thus we have,
\begin{eqnarray}
&& {\cal A}_n^{{\rm QED},L-s}(1_e^-,2_\gamma^+,\ldots,(n-1)_\gamma^+, n_e^+) 
   \nonumber \\
&& \hskip 2 cm = 
{i \over 2 } { (\sqrt{2} e)^n \over (4 \pi)^2} \sum_{\sigma \in S_{n-2}}
{\spa1.n \Sigma_{l=3}^{n-1}
       \sandmp1.{\Ksl_{\sigma(2)\cdots \sigma(l)} \ksl_{\sigma(l)}}.1
 \over \spa1.{\sigma(2)}\spa{\sigma(2)}.{\sigma(3)}\cdots
            \spa{\sigma(n-1)}.{n} \spa{n}.1} \,, 
\label{PhotonsA2}
\end{eqnarray}
where the sum over $\sigma$ runs over all permutations of legs
$2, 3, \ldots, n-1$, the subscript $e$ signifies an electron,
and the subscript $\gamma$ signifies a photon.
We have replaced  the QCD coupling $g$ by the QED coupling $\sqrt{2} e$, 
where the extra $\sqrt{2}$ is due to our normalization of the color 
matrices ($\Tr(T^a T^b) = \delta^{ab}$). This
result matches ${\cal A}_2$ as given in eq.~(77) of 
ref.~\cite{MahlonQED}, up to an overall factor of
$(-1)^{n+1}$.  A factor of $(-1)^{n}$ in this difference probably comes 
from an opposite sign convention for the polarization vector of a 
positive-helicity massless photon (or gluon).
The remaining sign may arise from an external-fermion sign convention.
(We use a convention compatible with supersymmetry, as described in
ref.~\cite{TwoQuarkThreeGluon}.)

Now consider Mahlon's ${\cal A}_1$ piece having closed fermion loops. 
The conversion of our primitive amplitudes with a closed fermion loop
to a QED amplitude is again given by a permutation sum over all gluon
legs to convert them to photons,
\begin{eqnarray}
&&
\hskip - 3 cm {\cal A}_n^{\rm QED,\ fermion\ loop}(1_e^-,
 2_\gamma^+,\ldots,(n-1)_\gamma^+, n_e^+)
\nonumber\\ 
&=& 
- { (-\sqrt{2} e)^n \over (4 \pi)^2}
\sum_{\sigma \in S_{n-2}}
 A_n^{s}(1_{\f}^-,n_\f^+, \sigma(2)^+, \ldots, \sigma(n-1)^+)\,,
\label{PhotonsA1}
\end{eqnarray}
where the permutations $\sigma$ are the same as in \eqn{PhotonsA2}.
The overall sign accounts for the sign difference between a scalar and
fermion in the loop. The factor of $(-1)^n$ takes into account
a minus sign between gluons emitted on the left side of the fermion
line, and those emitted on the right side.
Eq.~(68) of ref.~\cite{MahlonQED} gives ${\cal A}_1$.  If we multiply
that result by the same $(-1)^{n+1}$ normalization factor as for 
${\cal A}_2$, and convert it to our notation, we obtain,
\begin{eqnarray}
&&{\cal A}_n^{\rm QED,\ fermion\ loop}(1_e^-,2_\gamma^+,
\ldots,(n-1)_\gamma^+, n_e^+) 
  \nonumber \\
&& \hskip0cm = 
{2 i \over 3 } { (\sqrt{2} e)^n \over (4 \pi)^2} 
\sum_{\sigma \in S_{n-2}}
{ \sandmp1.{\ksl_{\sigma(2)} \ksl_{\sigma(3)} 
            \ksl_{\sigma(4)} \Ksl_{\sigma(2)\sigma(3)}}.1
  \over \spa{\sigma(2)}.{\sigma(3)} \spa{\sigma(3)}.{\sigma(4)} 
        \spa{\sigma(4)}.{\sigma(2)} \,
    \spa1.{\sigma(5)} \spa{\sigma(5)}.{\sigma(6)}
   \cdots \spa{\sigma(n-1)}.{n} }
  \nonumber \\
&& \hskip3.5cm
\times  
  {1 \over (k_{\sigma(2)}+k_{\sigma(3)}+k_{\sigma(4)})^2 } \,.
\label{MahlonPhotonsA1}
\end{eqnarray}
We have confirmed numerically that this result 
matches \eqn{PhotonsA1} through $n=9$.

We have also recovered the same QED amplitude~(\ref{MahlonPhotonsA1})
using primitive amplitudes with the positive-helicity
fermion at other locations in the ordering, by adjusting the
combinatoric factors appropriately.  For the class of $A_n^s$
primitive amplitudes where the positive-helicity fermion is in 
the $j^{\rm th}$ position, as in \eqn{alljn}, we first relabel the amplitude
so that the gluons still run from 2 to $n-1$, by letting 
$j\to n$, $j+1 \to j$, etc.  Then we sum over permutations
according to,
\begin{eqnarray}
&&
{\cal A}_n^{\rm QED,\ fermion\ loop}(1_e^-,
 2_\gamma^+,\ldots,(n-1)_\gamma^+, n_e^+)
\nonumber\\ 
&&= 
- { (-1)^j \over ( {n-5 \atop j-2} ) }
{ (-\sqrt{2} e)^n \over (4 \pi)^2}
\sum_{\sigma \in S_{n-2}}
 A_n^{s}(1_{\f}^-,\sigma(2)^+,\ldots,\sigma(j-1)^+,
                    n_f^+,\sigma(j)^+,\ldots,\sigma(n-1)^+)\,.~~~~
\label{jPhotonsA1}
\end{eqnarray}
The sign factor is $(-1)^j$ because now only $n-2-j$ gluons are 
emitted from the ``right'' side of the fermion line.  

The combinatoric
factor $( {n-5 \atop j-2} )$ can be explained simply if we assume 
that all QED contributions vanish whenever more than four photons attach 
to the massless electron loop --- one photon virtual, the rest real and 
of positive helicity.  This assumption is quite reasonable
because the multi-photon amplitudes 
${\cal A}^{\rm QED}(\gamma^\pm \gamma^+ \cdots \gamma^+)$
vanish for all $n>4$, for either sign of the first photon's 
helicity~\cite{MahlonQED}.
The contributions with fewer than four photons vanish by Furry's theorem,
and the vanishing of massless external leg corrections.
Then of the $n-2$ external gluons in the primitive amplitude,
precisely three must attach to the fermion loop, and $n-5$ to the external
fermion line.  The latter $n-5$ gluons are divided into $j-2$ on the left side
and $(n-5)-(j-2) = n-j-3$ on the right.  When we sum over all gluon
permutations, we overcount QED diagrams that differ only by a re-ordering
of photons on the left with respect to those on the right,
which preserves the order within the left set, and within the right set.
This overcount is $(n-5)!/(j-2)!/(n-j-3)! = ( {n-5 \atop j-2} )$.
Conversely, we can take the fact that \eqn{jPhotonsA1} works 
numerically (for all $j$ and $n\leq9$) as evidence in favor
of the vanishing of the one-off-shell, $(n-1)$-positive-helicity photon
amplitudes for $n>4$.

We also may use our primitive amplitudes to construct mixed QED/QCD 
amplitudes for an electron-positron pair plus $(n-2)$ 
positive-helicity gluons, and compare with Mahlon's earlier 
computation~\cite{Mahlon}.  In this case, the permutation
sums are a bit more involved. The mixed amplitudes are obtained using
two separate sums, which cancel contributions where a gluon is 
attached to the electron line, and which allow for all possible
orderings of the virtual photon attaching the $e^+e^-$ pair to the loop,
with respect to the color-ordered gluons.  We again let the negative-helicity
fermion be leg 1, but label the positive-helicity one by 2 (instead of $n$).

In the five-point case, for example, the appropriate sum for
the coefficient of the color trace $\Tr(T^{a_3} T^{a_4} T^{a_5})$ 
in ${\cal A}_5(1_e^-, 2_e^+, 3^+, 4^+, 5^+)$ is,
\begin{eqnarray}
 A_{5;1}^{\rm mixed}(1_e^-, 2_e^+, 3^+, 4^+, 5^+) 
 &=& -  { (\sqrt{2} e)^2 g^3 \over (4 \pi)^2} \sum_{\sigma \in Z_{3}} 
 \Bigl(A_n^{s}(1_\f^-, 2_\f^+, \sigma(3)^+, 
                              \sigma(4)^+, \sigma(5)^+) \nonumber \\
 && \hskip 3.5 cm \null
 + A_n^{s}(1_\f^-, \sigma(3)^+, 2_\f^+, \sigma(4)^+, \sigma(5)^+) \Bigr) \,,
\hskip 1cm 
\end{eqnarray}
where the sum runs over the cyclic permutations, $Z_3$, of the gluons
legs $\{3,4,5\}$. The unwanted diagrams appear in pairs in the
permutation sum, but with opposite signs due to the antisymmetry of
color-ordered vertices.  

More generally, the coefficient of the color structure
$\Tr(T^{a_3} T^{a_4} \ldots T^{a_n})$ 
in the $n$-point amplitude ${\cal A}_n(1_e^-, 2_e^+, 3^+,\ldots, n^+)$ 
is,
\begin{eqnarray}
&& A_{n;1}^{\rm mixed}(1_e^-, 2_e^+, 3^+,\ldots, n^+)  \nonumber \\
&& \hskip 0.3 cm \null 
= - {(\sqrt{2} e)^2 g^{n-2} \over (4 \pi)^2}
   \sum_{\sigma \in  Z_{n-2}} \sum_{j=2}^{n-2} 
    A_n^{s}(1_\f^-, \sigma(3)^+, \ldots, \sigma(j)^+, 
       2_f^+, \sigma(j+1)^+, \ldots, \sigma(n)^+) \,.
\hskip 1 cm 
\label{Mixed}
\end{eqnarray}
The permutation sum is over cyclic permutations, $Z_{n-2}$, of 
the $n-2$ gluon legs, labeled by $\{3,4,\ldots,n\}$.
The sum over $j$ is over the primitive amplitudes $A_n^s(j_\f^+)$ 
with the positive-helicity fermion in the $j^{\rm th}$ position.  
(The primitive amplitudes with fewer than two trailing gluons, 
$j=n-1$ or $n$, vanish trivially and have been dropped from the sum.)

Although Mahlon's corresponding formula, eq.~(53) of ref.~\cite{Mahlon},
contains some errors, it is not difficult to use his eqs.~(27) or~(33)
to rederive an expression for his form of the amplitude.
(The limit on the final sum in eq.~(33) should start at $\ell=2$, 
not $\ell= 1$.)  With our normalization conventions and labelling,
the result reads,
\begin{eqnarray}
A_{n;1}^{\rm mixed} (1_e^-,  2_e^+, 3^+,\ldots,n^+) 
&=&  - {i \over 3}  { (\sqrt{2} e)^2 g^{n-2}\over (4 \pi)^2} \, 
      {1 \over \spa3.4 \cdots \spa{(n-1)}.n \spa{n}.3 s_{12}} \nonumber\\
&&\null \times 
\sum_{\sigma \in Z_{n-2}}  \spb2.{\sigma(3)} \biggl(
  - \sandmp{\sigma(3)}.{\Ksl_{\! 12}\, \ksl_{\sigma(3)}}.1 
\label{MixedMahlon}\\
&& \null \hskip 2 cm 
  + 2 \sum_{l=4}^{n-1} \sandmp{\sigma(3)}.
        {\Ksl_{\sigma((l+1)\ldots n)}\, \ksl_{\sigma(l)}}.1 \biggr) \,.
 \nonumber 
\end{eqnarray}
The permutation sum again runs over cyclic permutations of 
the $n-2$ gluon legs, $\{3,4,\ldots, n\}$.
We have confirmed numerically through $n=16$ that our expression
(\ref{Mixed}) agrees with \eqn{MixedMahlon}.  This check is 
particularly useful because all primitive amplitudes 
$A_n^s(j_\f^+)$ enter into the permutation sum.

\subsection{Consistency of Various Shifts}

We can also make different choices for the shifted
momenta~(\ref{SpinorShift}) used to derive the recursion relations.
For many choices, the unreal poles found in the previous section lead
to extra terms with different soft-factor coefficients, which we again
determined empirically.  In other cases, the unreal poles are absent
entirely.  These latter choices give us an independent check on the
factors described above, because there are no correction factors to be
determined at all.  Only the ``naive'' terms, with the expected
single-power denominators, enter.  It turns out that we can make such
``clean'' choices for all amplitudes $A_n^{s}(j_\f^+)$ with $j>2$.

\begin{figure}[t]
\centerline{\epsfxsize 4.8 truein \epsfbox{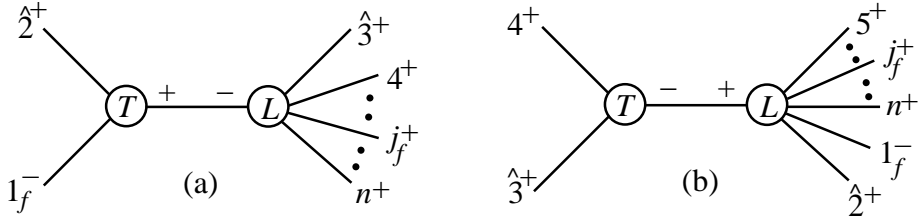}}
\caption{The two recursive diagrams for $A_n^{s}(j_\f^+)$ with $j>2$, 
using the shift in \eqn{SpinorShift32}.  Diagram (b) vanishes. 
Unreal poles do not appear in diagram (a). }
\label{An32sFigure}
\end{figure}

An especially nice choice with this property is the $(k,l)=(3,2)$ shift,
\begin{eqnarray}
\tlambda_3 &\rightarrow& \tlambda_3 - z\tlambda_2 \,,\nonumber\\
\lambda_2 &\rightarrow& \lambda_2 + z\lambda_3 \,. \label{SpinorShift32}
\end{eqnarray}
With this shift only the two diagrams displayed in \fig{An32sFigure}
are generated; other potential diagrams do not contribute because the
tree amplitude appearing in the factorization vanishes or because they
would contribute to $R$ type primitive amplitudes instead of the $L$
type under consideration.  Since the diagrams factorize onto
tree-level (not one-loop) three-point vertices,
they do not contain any unreal poles.  Very
neatly, diagram (b) vanishes because the three-point tree 
$A_3^{\tree}((-\Kh_{34})^-,\hat{3}^+,4^+)$ vanishes
with the shift (\ref{SpinorShift32}), leaving only diagram (a).
Furthermore, the evaluation of diagram (a) is rather simple, being
essentially the same as for MHV tree amplitudes.  For $A_n^s$,
inserting the value of the $(n-1)$-point amplitude from \eqn{alljn}
into this diagram immediately yields the corresponding $n$-point
amplitude, providing a simple verification of \eqn{alljn} for $j>2$,
with the $j=2$ boundary case as input.

%

For $j=2$ the shift~(\ref{SpinorShift32}) generates a surface term,
since the shifted amplitude does not vanish as $z\rightarrow
\infty$. However, the surface term can be taken into account easily
for checking a result, if not deriving it, as was
demonstrated for the all-plus $n$-gluon
amplitudes~\cite{OnShellRecurrenceI}.  There is no unreal-pole
contribution because the vertex 
$V_3^{\oneloop}(\hat 3^+, 4^+,(-\Kh_{34})^+)$ vanishes in the 
shifted kinematics.  We have checked that
this recursion relation is satisfied by the solution~(\ref{alljn}) for
$j=2$.

Similarly, for the case of $A_n^{L-s}(j_\f^+)$, with the same shift,
$(k,l) = (3,2)$, we also can construct a recursion relation with a
surface term, which is obeyed by \eqn{allnLminuss} for $j>2$, again
with input from the $j=2$ case.  (For $j=2$, this shift produces an
$A(z)$ which diverges as $z \rightarrow \infty$, so no useful recursion
relation is obtained.)

Numerous other choices of shifts are possible.  For many, if not all,
choices of shift variables, the contribution of an unreal pole will
have the form (in the $s_{ab}$ channel),
\begin{equation}
\pm{i\over s_{ab}}
 \, A_R^-  \,  V_3(\hat{a}^+,b^+,(-\Kh_{ab})^+)\, 
\, \Soft^\tree(b, (-\Kh_{ab})^-, \hat a)
\biggl(\sum_{i} c_i \, \Soft^\tree(u_i,\Kh_{ab}^+,v_i)\biggr),
\end{equation}
where the coefficients $c_i$ are either $1$ or $2$, $A_R$ is a tree
amplitude and $V_3$ is the three-positive helicity loop vertex.
(We can make the relative signs positive by interchanging
$u_i\leftrightarrow v_i$ if necessary.)  

In summary, in all shifts that we have checked, one can obtain consistent
results by appropriately adjusting the unreal pole contributions. For
$j>2$ we avoid the unreal poles with a suitable choice of shifts.
More stringent tests along these lines would require a first
principles derivation of the factors appearing in unreal poles.  We
defer this to future study.

\section{Conclusions}
\label{ConclusionSection}

In this paper, we have presented compact formul\ae{} for QCD
amplitudes with one quark pair, and $n-2$ gluons of identical helicity.
As explained in \sect{NotationSection}, these amplitudes are built out
of primitive amplitudes $A_n^{L-s}$ and $A_n^s$.  Our principal
results are \eqns{allnLminuss}{alljn}, for the all-$n$ forms of these
primitive amplitudes with positive-helicity gluons.  (The
corresponding primitive amplitudes with negative-helicity gluons can
be obtained by parity, implemented by spinor conjugation.)  
We also provide a new compact representation~(\ref{oneminusalln}) of 
the previously-obtained~\cite{Mahlon}
$n$-gluon amplitudes with a single negative helicity and the rest
positive. 

The corresponding tree-level quark-gluon amplitudes vanish, and
hence these amplitudes are both infrared- and ultraviolet-finite.
These were the last unknown finite loop amplitudes: formul\ae{} for
the finite $n$-gluon amplitudes~\cite{AllPlus,Mahlon} 
have been known for a while, and amplitudes with additional quark
pairs or higher loops are necessarily divergent in four dimensions.

Because the corresponding tree-level amplitudes vanish, the finite
one-loop amplitudes we computed do not enter next-to-leading order cross 
sections for jet production at hadron colliders.  However, they
do contribute at next-to-next-to-leading order.
The finite amplitudes also appear in factorization limits of the 
remaining, divergent one-loop QCD amplitudes.  Accordingly, their structure 
will likely play a role in understanding the latter amplitudes.

We constructed these amplitudes via on-shell recursion relations.
The construction of such relations relies on knowledge of
the factorization properties of amplitudes in {\it complex\/} momenta.
At tree level, these properties are determined by the factorization
properties in {\it real\/} momenta, which are known to be universal.
At loop level, this is no longer true.  Generic loop-level relations
differ from tree-level ones in having ``unreal
poles'' --- poles that are present for complex momenta, but absent for
real momenta.  In lieu of analogs of the standard real-momentum
factorization arguments~\cite{Neq4Oneloop,BernChalmers,OneloopSplit}
for this class of poles, we took a heuristic approach.  We empirically
determined the structure of terms in the recursion relations associated
with the unreal poles with the aid of known four- and five-point
amplitudes.  Then we applied the same structure to the case of 
additional external legs.  For most (but not all) of
the primitive amplitudes, we were able to find choices of complex
shifted momenta which avoid the unreal poles.  The agreement of the
results with these alternate shifts provides a strong consistency
check on the approach we took.  In addition, we verified that our
results satisfy the required collinear and multiparticle-pole
factorization forms (in real momenta).  A third and independent
stringent check comes from a comparison of certain QED and mixed
QCD/QED amplitudes computed by an entirely different method.

Unreal poles are an essential feature of the analytic structure of
loop amplitudes which deserves further study.  A first-principles
understanding of the extent of their universality and the structure of
factorization would be very important.  Such an understanding would
strengthen the use of loop-level recursion relations as a complement
to the unitarity-based method for performing loop calculations.


\section*{Acknowledgments}

We thank Academic Technology Services at UCLA for computer support. 



\appendix

\section{Multi-particle Factorization of $A_n^s(j_\f^+)$}
\label{MultiFactAppendix}

In this appendix we verify that the amplitudes~(\ref{alljn})
have the correct multi-particle factorization properties.  
As mentioned in \sect{OneLoopFact}, these helicity amplitudes 
only have multi-particle poles with gluonic intermediate states,
not fermionic ones.  
Given that the sum $S_2$ in \eqn{alljnS2} contains manifest
factors of $1/s_{l\cdots p}$, it is 
convenient to first consider the limit $K_{l\cdots p}^2 \to 0$ 
where $j < l$ and $l+1 < p < n$.   All the non-trivial multi-particle poles
are covered by this case, except for those with $p=n$, which
we shall discuss subsequently.

As $K_{l\cdots p}^2 \to 0$, we expect to find that
\begin{eqnarray}
&&
A_n^{s}(j_\f^+) \inlimit^{K_{l\cdots p}^2\rightarrow 0}
A_{n-p+l}^\tree(1_\f^-,2^+,\ldots,j_\f^+,\ldots,(l-1)^+,(-P)^-,(p+1)^+,\ldots,n^+)
\nonumber \\
&&\hskip30mm
\times {i \over K_{l\cdots p}^2 } \
A_{p-l+2}^{\oneloop}(P^+,l^+,\ldots,p^+) \,,
\label{lppole}
\end{eqnarray}
where $P=-K_{l\cdots p}$, and $A_{p-l+2}^\oneloop$ is the all-plus 
pure-glue amplitude~(\ref{OneLoopAllPlusAmplitude}).
Let $H_{p-l+2}(l,p)$ denote the numerator factor $H_n$ 
defined in \eqn{Hndef}, after the external momenta are
relabeled to correspond to 
$A_{p-l+2}^\oneloop(P^+,l^+,\ldots,p^+)$.
Using also \eqn{ffmhvtree} for the quark-containing tree amplitudes,
we expect the behavior
\begin{eqnarray}
&&
A_n^{s}(j_\f^+) \inlimit^{K_{l\cdots p}^2\rightarrow 0}
{i \over 3} { {\spa1.{P}}^3 \spa{j}.{P} 
  \over \spa1.2 \cdots \spa{(l-2)}.{(l-1)} \spa{(l-1)}.{P} 
         \spa{P}.{(p+1)} \cdots \spa{n}.1 }
\nonumber\\
&& \hskip30mm
\times 
{1\over s_{l\cdots p} } \,
 { H_{p-l+2}(l,p)
  \over \spa{P}.{l} \spa{l}.{(l+1)} \cdots \spa{(p-1)}.{p} \spa{p}.{P} }
\,.
\label{lppole2}
\end{eqnarray}
Now examine the term labeled by $l$ and $p$ in \eqn{alljnS2} for $S_2$,
in the multi-particle factorization limit.  Using 
$K_{2\cdots(l-1)} = - k_1 + P - K_{(p+1)\cdots n}$, we have
\begin{eqnarray}
 &&S_2(l,p) \inlimit^{K_{l\cdots p}^2\rightarrow 0} 
 - { \spa{(l-1)}.{l} 
   \over \spab1.{\Ksl_{(p+1)\cdots n}}.{P} \spa{P}.{(l-1)}
         \spab1.{\Ksl_{(p+1)\cdots n}}.{P} \spa{P}.{l} }
\nonumber \\
&& \hskip30mm
\times 
 {  \spa{p}.{(p+1)} 
   \over \spab1.{\Ksl_{(p+1)\cdots n}}.{P} \spa{P}.{p}
         \spab1.{\Ksl_{(p+1)\cdots n}}.{P} \spa{P}.{(p+1)} }
\nonumber \\
&& \hskip30mm
\times
   { { \spa{1}.{P}}^2 {\spab1.{\Ksl_{(p+1)\cdots n}}.{P}}^2
     \spa{j}.{P} \spab1.{\Ksl_{(p+1)\cdots n}}.{P}  }
\nonumber \\
&& \hskip30mm
\times
 { \spa1.{P} \spab1.{\Ksl_{(p+1)\cdots n} [ {\cal F}(l,p) ]^2}.{P}
    \over s_{l\cdots p} } \,,
\label{S2limit}
\end{eqnarray}
where we also used the fact that 
\begin{equation}
\sandpm{X}.{[ {\cal F}(l,p) ]^2}.{X} = 0
\label{XFX}
\end{equation}
for any spinor, or spinor string, $X$. 
\Eqn{XFX} follows from a more general result,
\begin{equation}
\sandpm{X}.{[ {\cal F}(l,p) ]^2}.{Y} 
= - \sandpm{Y}.{[ {\hat{\cal F}}(l,p) ]^2}.{X} 
= - \sandpm{Y}.{[ {\cal F}(l,p) ]^2}.{X}  \,,
\label{XFY}
\end{equation}
where the lengths of strings $X,Y$ are the same mod 2 (otherwise
there is an additional sign in reversing them).  Here we have defined
the reversal of ${\cal F}(l,p)$,
\begin{equation}
{\hat{\cal F}}(l,p) \equiv \sum_{j=l}^{p-1} \sum_{i=j+1}^{p} \ksl_i \ksl_j \, 
\label{hatFlpdef}
\end{equation}
so that
\begin{equation}
{\cal F}(l,p) + {\hat{\cal F}}(l,p) = \Bigl( \sum_{i=l}^p k_i \Bigr)^2 =
K_{l\cdots p}^2 = 0
\end{equation}
in the factorization limit.
Comparing the limiting behavior~(\ref{S2limit}) of $S_2$ with the
expectation~(\ref{lppole2}), and cancelling various factors of
$\spab1.{\Ksl_{(p+1)\cdots n}}.{P}$, we see that the limit will
be correct if we can show that 
\begin{equation}
\spab1.{\Ksl_{(p+1)\cdots n} [ {\cal F}(l,p) ]^2}.{P}
 = - \spab1.{\Ksl_{(p+1)\cdots n}}.{P} 
   H_{p-l+2}(l,p) \,.
\label{FFtoshow}
\end{equation}
The Schouten identity,
\be
  \spb{a}.{b} \spb{c}.{d} 
= \spb{a}.{c} \spb{b}.{d} + \spb{a}.{d} \spb{c}.{b} \,,
\label{Schoutenb}
\ee
together with \eqn{XFY}, imply that 
\begin{eqnarray}
\spab1.{\Ksl_{(p+1)\cdots n} [ {\cal F}(l,p) ]^2}.{P}
 &=& \spab1.{\Ksl_{(p+1)\cdots n}}.{P} \Tr_{-} [ {\cal F}(l,p) ]^2
\nonumber \\
&& \hskip2mm
 - \spab1.{\Ksl_{(p+1)\cdots n} [ {\cal F}(l,p) ]^2}.{P} \,,
\end{eqnarray}
or
\begin{equation}
\spab1.{\Ksl_{(p+1)\cdots n} [ {\cal F}(l,p) ]^2}.{P}
 = {1\over2} \spab1.{\Ksl_{(p+1)\cdots n}}.{P} \Tr_{-} [ {\cal F}(l,p) ]^2 \,.
\label{FFspinoreq}
\end{equation}
So we just need to show that
\begin{equation}
{1\over2} \Tr_{-} [ {\cal F}(l,p) ]^2
= - H_{p-l+2}(l,p) \,.
\label{FFtoshow2}
\end{equation}

We first use momentum conservation to remove the terms
in $H_{p-l+2}(l,p)$ (see \eqn{Hndef}) which contain the 
massless leg $P$:
\begin{eqnarray}
 - H_{p-l+2}(l,p) &=&  \sum_{l\leq i_1<i_2<i_3<i_4\leq p} 
  \Tr_{-}\Bigl[\ksl_{i_1}\ksl_{i_2}\ksl_{i_3}\ksl_{i_4}\Bigr] 
  - \sum_{l\leq i_1<i_2<i_3\leq p} \sum_{i_4=l}^{p} 
   \Tr_{-}\Bigl[\ksl_{i_1}\ksl_{i_2}\ksl_{i_3}\ksl_{i_4}\Bigr] 
\nonumber\\
 &=& - \sum_{l\leq i_1<i_2<i_3 \leq p} \sum_{i_4=l}^{i_3-1} 
   \Tr_{-}\Bigl[\ksl_{i_1}\ksl_{i_2}\ksl_{i_3}\ksl_{i_4}\Bigr] \,.
\label{newHnsum}
\end{eqnarray}
On the other hand, the left-hand side of \eqn{FFtoshow2} can be 
rewritten using the Schouten identity~(\ref{Schoutenb}) as,
\begin{eqnarray}
{1\over2} \Tr_{-} [ {\cal F}(l,p) ]^2
&=& {1\over2} \sum_{l \leq i_1 < i_2 \leq p} \, \sum_{l \leq i_3 < i_4 \leq p} 
  \spa{i_1}.{i_2} \spb{i_2}.{i_3} \spa{i_3}.{i_4} \spb{i_4}.{i_1}
\nonumber\\
&=& {1\over2} \sum_{l \leq i_1 < i_2 \leq p} \, 
           \sum_{l \leq i_3 < i_4 \leq p} 
  \Bigl( - \spa{i_1}.{i_2} \spb{i_2}.{i_4} \spa{i_4}.{i_3} \spb{i_3}.{i_1}
         + \spa{i_1}.{i_2} \spb{i_2}.{i_1} \spa{i_3}.{i_4} \spb{i_4}.{i_3}
  \Bigr)
\nonumber\\
&=& - {1\over2} \sum_{l \leq i_1 < i_2 \leq p} \, 
            \sum_{l \leq i_4 < i_3 \leq p} 
    \spa{i_1}.{i_2} \spb{i_2}.{i_3} \spa{i_3}.{i_4} \spb{i_4}.{i_1} \,.
\label{FFid1}
\end{eqnarray}
The last term in the second line vanishes using $K_{l\cdots p}^2 = 0$.
On the right-hand side of \eqn{FFid1} we split the sum over $i_2$ and
$i_3$ into two pieces, one with $i_2 < i_3$ and the second with $i_2 > i_3$.
The first sum has $i_1 < i_2 < i_3$ and $i_4 < i_3$. 
It manifestly agrees with the right-hand side of \eqn{FFtoshow2},
as given in \eqn{newHnsum},
up to an overall factor of 1/2.  The second sum
has $i_4 < i_3 < i_2$ and $i_1 < i_2$.  
Relabelling the indices $i_1 \lr i_4$, $i_2 \lr i_3$ and reversing the
order of the spinor string, we see that the second sum is precisely
equal to the first sum.  Adding the first and second sums together
proves the identity~(\ref{FFtoshow2}), which establishes the proper
multi-particle factorization behavior~(\ref{lppole}) of the 
amplitudes $A_n^{s}(j_\f^+)$ for $p < n$.

Now consider the remaining cases where $K_{l\cdots n}^2 \to 0$. 
These multi-particle poles are in fact the ones appearing in the
$l^{\rm th}$ term in the recursive construction~(\ref{SFullRecurrence}).
In \eqn{alljn}, the relevant $1/s_{l\cdots n}$ 
poles are hidden in the terms 
with $p=n-1$ in \eqn{alljnS2} for $S_2$.  They can be found 
in the factor
\begin{equation}
\sandmp1.{\Ksl_{2\cdots (l-1)} \Ksl_{l\cdots p}}.{(p+1)}
= \sandmp1.{\Ksl_{2\cdots (l-1)} \Ksl_{l\cdots n}}.{n} 
= - \spa1.{n} s_{l\cdots n} \,,
\end{equation}
after using momentum conservation, 
$K_{2\cdots (l-1)} = - k_1 - K_{l\cdots n}$.
Note also that in this limit, with $P = - K_{l\cdots n}$,
we have 
\begin{eqnarray}
K_{(p+1)\cdots n} &=& k_n \,,
\\
\langle 1^- | \Ksl_{2\cdots (l-1)} &=& \spa1.{P} \langle P^+ | \,,
\\
s_{l\cdots p} &=& s_{l\cdots (n-1)} = (- P - k_n)^2 
= \spa{n}.{P} \spb{P}.{n} \,. 
\end{eqnarray}
Using these relations, the limiting behavior of the $(l,p=n-1)$ term
in $S_2$, as $K_{l\cdots n}^2 \to 0$, is
\begin{eqnarray}
 &&S_2(l,n-1) \inlimit^{K_{l\cdots n}^2\rightarrow 0} 
  { \spa{(l-1)}.{l} 
   \over {\spa1.{n}}^2 {\spb{n}.{P}}^2 \spa{P}.{(l-1)} \spa{P}.{l} }
\nonumber \\
&& \hskip30mm
\times 
 {  \spa{(n-1)}.{n} 
   \over \spa1.{P} \spb{P}.{n} \spa{n}.{(n-1)} \spa1.{n} s_{l\cdots n} }
\nonumber \\
&& \hskip30mm
\times (-1) \times
   { \spa{1}.{P}}^2 {\spb{P}.{n}}^2 {\spa{n}.1}^2 
     \spa{j}.{P} \spb{P}.{n} \spa{n}.1
\nonumber \\
&& \hskip30mm
\times
 { \spa1.{P} \sandpm{P}.{[ {\cal F}(l,n-1) ]^2}.{n} \spa{n}.1
    \over \spa{n}.{P} \spb{P}.{n} } \,,
\nonumber \\
&& \hskip25mm
= - { {\spa1.{P}}^3 \spa{j}.{P} \over s_{l\cdots n} }
 { \spa{(l-1)}.{l} \over \spa{(l-1)}.{P} \spa{P}.{l} }
 { \spa{n}.1 \over \spa{n}.{P} \spa{P}.1 }
 { \sandpm{n}.{[ \hat{\cal F}(l,n-1) ]^2}.{P} \over \spb{n}.{P} }  \,.
\nonumber\\
&&~~
\label{S2speciallimit}
\end{eqnarray}
If the second argument of $\hat{\cal F}(l,n-1)$ in \eqn{S2speciallimit} 
were $n$ instead of $(n-1)$, we would be done, as we could then use
the same logic as in the case $p < n$ treated earlier.
But first we have to show that 
\begin{equation}
\sandpm{n}.{[ \hat{\cal F}(l,n-1) ]^2}.{P} 
= \sandpm{n}.{[ \hat{\cal F}(l,n) ]^2}.{P} \,.
\label{Fextend} 
\end{equation}
From the definition~(\ref{hatFlpdef}), 
$\hat{\cal F}(l,n) - \hat{\cal F}(l,n-1) = \ksl_n \sum_{j=l}^{n-1} \ksl_j$.
The first $\hat{\cal F}(l,n-1)$ on the left-hand side
of \eqn{Fextend} can be replaced by $\hat{\cal F}(l,n)$, because
$\spb{n}.{n} = 0$.  The second one can be replaced as well,
because the difference is proportional to 
$\sum_{j=l}^{n} \spab{n}.{j}.{P} = - \spa{n}.{P} \spb{P}.{P} = 0$.
This verifies the factorization behavior as $K_{l\cdots n}^2 \to 0$.


\end{document}